\documentclass[a4paper,fleqn]{cas-dc}

\usepackage[authoryear]{natbib}
\usepackage{lineno,hyperref}
\usepackage{subcaption}
\usepackage{tabularx,booktabs}
\usepackage{bm}
\def\tsc#1{\csdef{#1}{\textsc{\lowercase{#1}}\xspace}}
\tsc{WGM}
\tsc{QE}

\begin{document}
\let\WriteBookmarks\relax
\def\floatpagepagefraction{1}
\def\textpagefraction{.001}
\shorttitle{Conceptual Study and Performance Analysis of Tandem Dual-Antenna
Spaceborne SAR Interferometry}    

\shortauthors{F. Hu et~al}  

\title [mode = title]{Conceptual Study and Performance Analysis of Tandem Dual-Antenna Spaceborne SAR Interferometry}  



%

\author[1]{Fengming Hu}[orcid=0000-0001-6911-1073]
\cormark[1]
\ead{fm_hu@fudan.edu.cn}

\author[1]{Feng Xu}

\author[2,3]{Robert Wang}

\author[4]{Xiaolan Qiu}

\author[4]{Chibiao Ding}

\author[1]{Yaqiu Jin}

\address[1]{the Key Laboratory for Information Science of Electromagnetic Waves (MoE), Fudan University, Shanghai 200433, China}

\address[2]{the Department of Space Microwave Remote Sensing System, Aerospace Information Research Institute, Chinese Academy of Sciences, Beijing 100190, China}

\address[3]{the School of Electronic, Electrical and Communication Engineering, University of Chinese Academy of Sciences, Beijing 101408, China}

\address[4]{the National Key Laboratory of Science and Technology on Microwave Imaging, Aerospace Information Research Institute, Chinese Academy of Sciences, Beijing 100094, China}















\begin{abstract}
Multi-baseline synthetic aperture radar interferometry (MB-InSAR), capable of mapping 3D surface model with high precision, is able to overcome the ill-posed problem in the single-baseline InSAR by use of the baseline diversity.
Single-pass MB acquisition with the advantages of high coherence and simple phase components has a more practical capability in 3D reconstruction than conventional repeat-pass MB acquisition. Using an asymptotic 3D phase unwrapping (PU), it is possible to get a reliable 3D reconstruction using very sparse acquisitions but the interferograms should follow the optimal baseline design. However, current spaceborne SAR system doesn't satisfy this principle, inducing more difficulties in practical phase unwrapping.
In this article, a new concept of Tandem Dual-Antenna SAR Interferometry (TDA-InSAR) system for acquiring optimal single-pass multi-baseline (MB) interferograms is proposed.
Two indicators, i.e., expected relative height precision and successful phase unwrapping rate, are selected to optimize the system parameters.
Additionally, the performances of various baseline configurations in typical scenarios and the impact of different error sources are investigated correspondingly. 
The simulation-based demonstrations shows that the proposed TDA-InSAR enables the fast 3D reconstruction in a single flight and optimal MB acquisitions for asymptotic 3D phase unwrapping. 
\end{abstract}


\begin{highlights}
\item 
A new TDA-InSAR is proposed which is tailored to get the specified optimal MB interferograms for asymptotic 3D PU algorithm, achieving fast 3D reconstruction.

\item Performances of different baseline configurations and the impact of different error sources are systematically investigated. 

\item Simulation-based performance evaluation is conducted, indicating that one example configuration of the proposed system can achieve a 3D reconstruction with a relative height precision of 0.3 m in built-up or man-made objects and that of 1.7 m in vegetation canopies. 
\end{highlights}


\maketitle


\section{Introduction}
%
%
%
%
{I}{nterferometric} synthetic aperture radar (InSAR) is a useful tool to get high-resolution topographic maps of terrain~\cite{bamler98,hanssen01}. 
The key part of InSAR processing is the phase unwrapping (PU).
The problem of 2D PU is ill-posed, which has to be solved under the assumption that the absolute phase difference between neighboring pixels is less than $\pi$, denoted as phase continuity (PC) assumption~\cite{ghiglia98}. This assumption is only valid for natural surface or interfergram with very short baseline. Nevertheless, using multi-baseline (MB) interferograms is proven to be an efficient way to overcome the PC assumption~\cite{ferretti99,Gini2005}.  

There are three types of MB interferograms according to the data acquisition mode, i.e., repeat-pass,
single-pass multi-antenna,
and
single-pass multi-satellite MB interferograms. 
{\color{black}Most spaceborne SAR systems obtain repeat-pass MB interferograms, which contains various error sources, such as atmospheric delay, orbit trend, deformation~\cite{zebker92}.}
The key part of the conventional 3D PU is separating different phase components based on their spatio-temporal characteristic~\cite{leijen14,Kampes2004,2007hooper3DPHU,shanker2007}. 
In case of persistent scatterer interferometry (PSI)~\cite{ferretti00,Fengming2019}, prior deformation model is used to decrease the phase gradient and the 3D data stack is unwrapped first in the baseline domain and then in the spatial domain.
In case of Small BAseline Subset (SBAS)~\cite{berardino02,mora03}, only interferograms with a short baseline are used to achieve a model-free 1D baseline PU. 
{\color{black}There are two limitations for the repeat-pass MB interferograms. First, a number of interferograms are required to improve the reliability, leading to a long acquisition period. Secondly, temporal decorrelation in vegetation canopies would decrease the coherence of the interferograms, which significantly affect the height precision.}
\begin{table*}
\caption{Comparison of MB interferograms obtained by different acquisition mode}
\begin{tabularx}{\linewidth}{XXXXX}
\toprule
\multirow{2}{*}{Acquisition mode} & Repeat-pass & \multicolumn{2}{c}{Single-pass} & TDA-InSAR\\
 & & multi-antenna &  multi-satellite & \\
\midrule
Satellite formulation & Single-satellite & Single-satellite & Multi-satellite & Tandem-satellite\\
Antenna configuration & Single-antenna & Multi-antenna & Single-antenna & Dual-antenna\\
Maximal baseline length & Long & Short & Long & Long\\ 
Optimal baseline design & No & Yes & No & Yes \\
Acquisition period & Long & Short & Short & Short \\ 
Coherence & Low & High & High & High \\
Working mode & mono-static & bi-static & bi-static & bi-static/mono-static \\
Error sources & atmospheric delay, orbit error, deformation, system noise & system noise & orbit error, system noise & orbit error, system noise, synchronization error \\
\bottomrule
\end{tabularx}
\label{tab:01}
\end{table*}

{\color{black}The single-pass MB interferograms suffer similar degrees of atmospheric effects, but the variation of the atmospheric delay is rather limited, which can be neglected in the practical process. Thus, the second and third types of MB interferograms can be directly unwrapped.}
There are three types MB PU methods
, i.e., the Chinese remainder theorem (CRT)-based method, projection method, and baseline linear combination (BLC) method~\cite{Xu1994}. 
The CRT-based methods~\cite{Xia2007,Wang2010TSP,Yuan2013} are widely used due to its robustness.
The cluster-analysis (CA)-based MB PU method {\color{black}extends the CRT-based methods by combining the statistical information and the spatial distribution of the intercept values~\cite{2010Yu,Liu2014}, which} enables a more reliable PU. However, rapid phase change and strong noise may decrease the reliability of the identified intercept clusters and previous CRT-based PU methods only focus on the local statistical information for the optimization. {\color{black}The phase difference based algorithms, such as maximum likelihood method~\cite{Fornaro2005} and Two-Stage Programming Approach (TSPA)~\cite{2016Robust} improve the noise robustness from the perspective of global optimization.}


Optimal baseline design plays an important part on the performance of all MB PU methods, especially for very sparse acquisitions. 
In~\cite{Yu2019OBD}, a nonlinear mixed-integer programming (NIP) criteria provides a credible lower bound of the baseline, and the followed closed-form robust CRT in~\cite{Yuan2020} gives a meaningful upper bound by considering the ambiguity height. Both indicate that 
increasing the number of acquisitions can not improve the height precision significantly if the longest baseline is limited. Using the accuracy information-theoretical assessment, research in~\cite{Ferraiuolo2009} shows that the final height precision only depends on the longest baseline length of the MB interferograms. 
For a very sparse acquisition, such as three or four single-pass SAR images, an asymptotic 3D PU is developed to achieve a robust 3D reconstruction~\cite{Hu2022Asymptotic}. Following a 2D (space) $+$ 1D (baseline) PU framework~\cite{Thompson1999}, it provides the optimal bounds for baseline design. 

{\color{black}The spaceborne single-pass SAR system, Shuttle Radar Topography Mission (SRTM) successfully get the global digital elevation model within $-56^\circ$ and $60^\circ$ latitude, showing that the single-pass dual-antenna interferogram has a good coherence~\cite{SRTM2003}. 
Since longer perpendicular baseline leads to better height precision, the SRTM uses double antennas with a 60-m cross-track separation and provides a height precision in the order of 10 m. 
Although the multi-antenna interferograms have the advantages of accurate baseline measurement and unique phase component corresponding to the ground elevation~\cite{Ding2019}. 
The short baseline length of the single-pass dual-antenna interferogram limits its practical application. 

The TanDEM-X mission consists of two satellites, which obtains the single-pass dual-satellite interferogram using a bi-static mode. This bi-static SAR 
with a longer baseline configuration provides a height precision of 2 m but requires strict synchronization between the satellites~\cite{Krieger2007}. Additionally, the dual-satellite interferograms shows a good coherence in vegetation canopies area, which is widely used in the tropical-forest biomass estimation~\cite{Torano2016}. 
The following LuTan-1 mission follows the similar design as the TanDEM-X but adopts the L-band to achieve a better performance in biomass inversion~\cite{Jin2020}. 
During the global coverage of the TanDEM-X, the height ambiguity was set to 45 m in the first global acquisition to avoid phase unwrapping errors and then that was set to 35 m in the second global acquisition to improve the height precision~\cite{Zink2014}. Inappropriate height ambiguity will lead to phase unwrapping errors, especially in urban area. To improve the reliability of the PU, the adjustment of the baseline is necessary but this will reduces the timeliness of the data acquisition. 

Additionally, {\color{black}multi-satellite SAR interferogram will suffer from orbit error due to the inaccuracies of the orbit parameters. Since the orbit inaccuracies are correlated in time, the orbit error often result in a spatially correlated phase trend in the interferometric phase, which will bias the estimated parameters. Since the illumination time for a LEO SAR is only few seconds, this phase trend can be estimated jointly with other parameters of interest using the plane function~\cite{Zhang2014} or the nonlinear model~\cite{Liu2016,Baehr2012}. }
Note that the orbit error can be compensated only if the multi-satellite interferogram is successfully unwrapped. 

Future high-resolution wide-swath (HRWS) mission will cooperate with three MirrorSAR satellites to acquire MB interferograms in a single flight. This mission has a better performance in height precision and timeliness than previous TanDEM-X mission by using the baseline diversity~\cite{Mittermayer2022}. 
However, the orbit uncertainty of the small satellite will affect optimal baseline design and induce additional orbit error. Such single-pass multi-satellite MB interferograms don't satisfy the requirements of the asymptotic 3D PU. Thus, conceptual study of new SAR system based on the optimal baseline design is necessary, which is the foundation of the work presented in this paper.  
}

To overcome the limitations of these conventional InSAR modes, a new spaceborne InSAR configuration named Tandem Dual-Antenna SAR Interferometry (TDA-InSAR) is proposed, aiming to acquire the {\color{black}specified} MB interferograms for the asymptotic 3D PU~\cite{Hu2022Asymptotic}. The comparison of MB interferograms obtained by different acquisition modes is shown in Table.~\ref{tab:01}.
It shows that the proposed TDA-InSAR combines the advantages of both multi-antenna and multi-satellite interferograms and thus provides the best baseline design. 

The paper is organized as follows. We introduce the concept of the TDA-InSAR and briefly review the main approach of the asymptotic 3D PU in section II. Investigation of the optimal baseline design using different baseline configuration is presented in Section III, followed by the simulation-based performance evaluation and analysis of various error sources in Section IV. The conclusions are presented in Section V.

\section{Concept of TDA-InSAR}

\subsection{Basic Concept}

2D radar image often suffers from severe geometric distortion, increasing the difficulty of target recognition. {\color{black}The key issue is estimating the height of all scaterers, }which can be well solved by using the MB interferograms. 

The phase components of the repeat-pass MB interferograms include flat earth effect, height, atmospheric delay, orbit error, deformation and system noise, which can be written as follows
\begin{equation}
\varphi = \textrm{wrap}\{\varphi_{\textrm{flat}} + \varphi_{\textrm{h}} + \varphi_{\textrm{atmos}} + \varphi_{\textrm{orb}} + \varphi_{\textrm{def}} + \varphi_{\textrm{noise}} + 2\pi n\},
\label{eq:phi}
\end{equation}
{where \mbox{$n\in \mathbb{Z}$} denotes the integer phase ambiguity.} In order to get the height estimation with a high precision, it is necessary to model different phase components appropriately. 
{\color{black}Because the atmospheric signal, orbital phase trend and deformation show a strong correlation in space, leading to a bias in the estimated height.}
{\color{black}The spatio-temporal characteristics of the error sources are used to separate the difference phase contributions. First, the deformation is assumed to be correlated in time and a prior deformation model is used to unwrap the MB interferograms. Then, the spatial trend due to the orbital inaccuracies, together with a trend in the atmospheric signal can be estimated using the unwrapped phase. 
Supposing the acquisitions are taken few hours apart, the atmospheric delay is uncorrelated in time but the the unmodelled deformation is assumed to be temporally correlated. A temporal low-pass filter can be applied to isolate the atmospheric delay from the unmodelled deformation. 
Following this way, it is possible to remove most of the atmospheric delay and orbital phase trend in the original phase. Note that a number of acquisitions is required to improve the reliability. The conventional single-beam SAR lacks the timeliness,}
which is not suitable for the specified applications with the near real-time requirement, such as target recognition. 

\begin{figure}[htbp]
     \centering
     \includegraphics[width=0.44\textwidth]{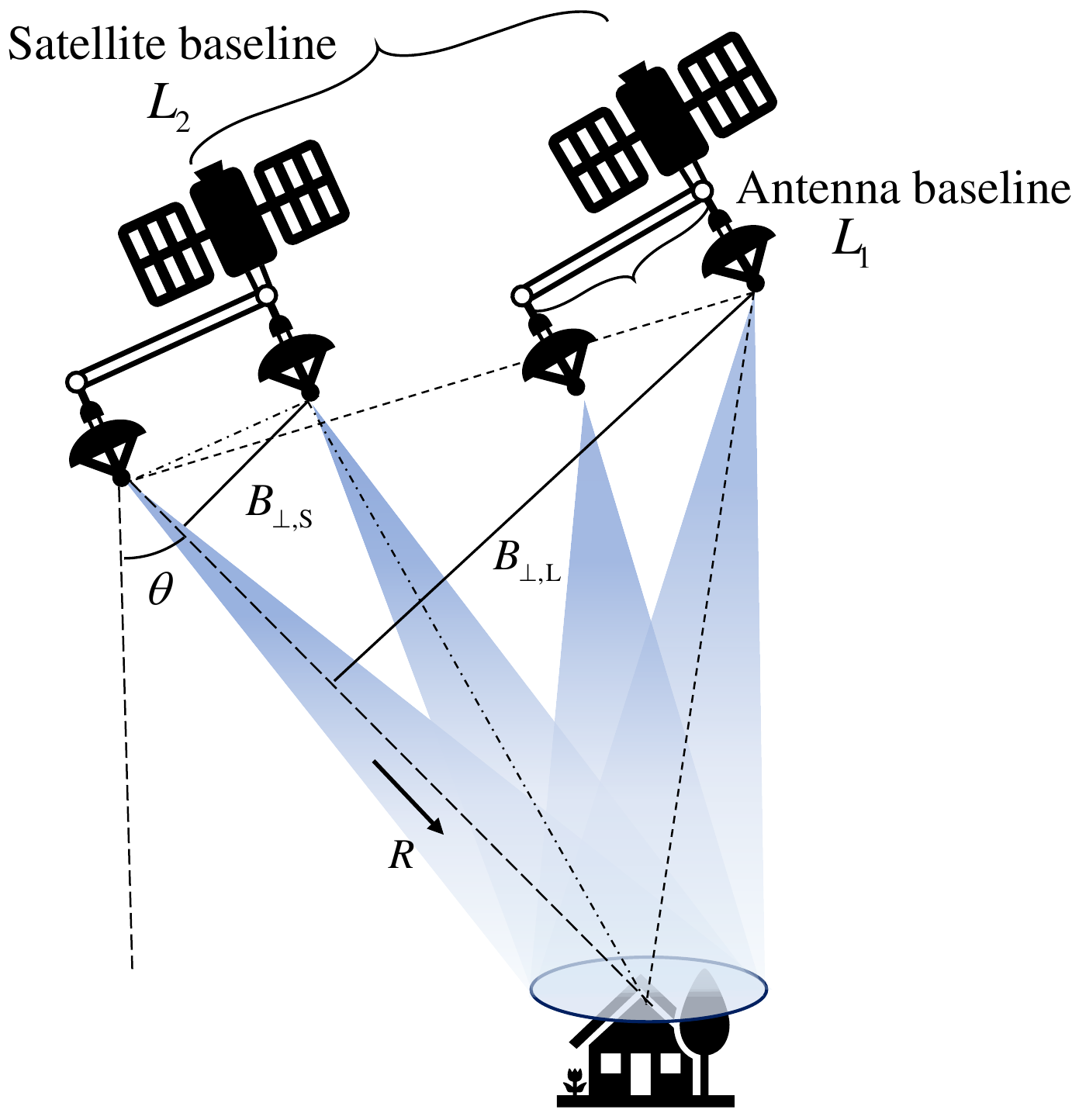}
     \caption{ TDA-InSAR imaging geometry. This system is able to acquires three independent interferograms in a single flight.}
     \label{fig:DDAS}
\end{figure}

Considering the single-pass MB interferograms, the atmospheric delay part in Eq.~(\ref{eq:phi}) can be neglected. Previous research shows that the asymptotic 3D PU enables a reliable 3D reconstruction with very sparse acquisitions. But the  
optimal single-pass MB interferograms should meet the following four conditions to achieve a good performance. 

1) There is a real or pseudo {\color{black}short baseline interferogram (SBI)}, satisfying the PC assumption. 

2) The phase components of SBI only include height and system noise. 

3) The ratio of the baseline length between SBI and {\color{black}long baseline interferogram (LBI)} satisfies the SU criteria. {\color{black}If not, medium baseline interferogram (MBI) should be used to improve the probability of successful unwrapping. }

4) The successful unwrapped LBI can achieve the expected height precision.


To meet the requirements of the optimal baseline design and decrease the impact of the orbit error, the proposed TDA-InSAR consists of two dual-antenna satellites, which acquires both dual-antenna and {\color{black}dual-satellite} interferograms in a single flight. 
The single-pass dual-antenna interferogram has the advantages of short baseline length, high coherence and independence of orbit error, which can be a good SBI. 
On the contrary, the single-pass dual-satellite interferogram enable a longer baseline, which can be a good LBI or MBI. Note that the dual-satellite interferogram can obtained by either bi-static or mono-static mode. The bi-static interferogram will contain the orbit error while the mono-static interferogram will include both orbit error and atmospheric delay.   

Fig.~\ref{fig:DDAS} is the simplified sketch of the TDA-InSAR. The main parameters of this system are antenna baseline $L_1$ and satellite baseline $L_2$. The antenna baseline should be as short as possible to decrease the difficulties in the hardware design. Thus, the optimal satellite baseline is expected to be in a flexible interval to suppress the uncertainty of the orbit error. Such system can work in either bistatic or mono-static mode. 
In case of the bistatic mode, the bistatic measurement will include additional phase errors in azimuth slow time~\cite{Krieger2007} because of the relative frequency deviation and phase noise between different radar instruments. Such phase error can be eliminated using either echo-domain~\cite{He2012} or image-domain algorithm~\cite{Krieger2007}, which is not investigated in this work.

In the following section, 
we review the asymptotic 3D PU algorithm 
and introduce the optimal baseline design for the TDA-InSAR. A modified 3D reconstruction approach with respect to the orbit error compensation is developed using the unwrapped phase.

\subsection{Asymptotic 3D PU for TDA-InSAR}
In a single flight, 
the proposed TDA-InSAR acquires three independent interferograms including both dual-antenna and {\color{black}dual-satellite} interferograms. 
{\color{black}
The dual-antenna interferogram is used as a SBI in the asymptotic 3D PU, providing the initial height estimation. The dual-satellite interferogram with the longest baseline is a LBI. Furthermore, other dual-satellite interferograms are used as the MBI, improving the reliability of the ambiguity estimation. In the following description, the subscripts $\textrm{S}$, $\textrm{M}$ and $\textrm{L}$ denote the parameters of SBI, MBI and LBI, respectively.}


Using the conventional InSAR process, the flat earth effect can be well determined using the parameters of the SAR system. After removing this phase, 3D PU can be applied to the flattened phase.
The spatial PU is the integration of the phase gradient in the spatial domain. Its reliability depends on the selection of the phase difference. However, whether an interferogram satisfies the PC assumption can not be evaluated using the single baseline interferogram. Alternatively, using the smoothing criteria with a pair of interferogram~\cite{Hu2022Asymptotic}, the reliability of the spatial PU can be improved. 

{\color{black}Since the dual-antenna interferogram always satisfies the PC assumption, the spatial PU can be conducted directly and the initial height estimation can be obtained as follows~\cite{hanssen01}
\begin{equation}
\color{black}
\hat{h}_\textrm{S} = \frac{\lambda}{4 \pi} \frac{R \sin \theta}{B_{\perp,\textrm{S}}}
\phi_{\textrm{S}}. 
\label{eq:h_s}
\end{equation}
where $\phi_{\textrm{S}}$ denotes the unwrapped phase of SBI. $B_{\perp}$ denotes the perpendicular baseline. $R$ is the slant range. $\theta$ is the incident angle and $\lambda$ is the radar wavelength.  }

{\color{black}Using the initial height,}
the pseudo unwrapped phase of MBI and the corresponding phase ambiguity can be estimated as follows
\begin{equation}\color{black}
\phi_{\textrm{pseudo},\textrm{M}}= \frac{4 \pi}{\lambda} \frac{B_{\perp,\textrm{M}}}{R \sin \theta}\hat{h}_\textrm{S} 
 = \frac{B_{\perp,\textrm{M}}}
{B_{\perp,\textrm{S}}} \phi_\textrm{S},
\end{equation}

\begin{equation}
\hat{n}_{\textrm{M},0} = \textrm{round} \left\{\frac{\phi_{\textrm{pseudo},\textrm{M}} - \textrm{wrap}\{\varphi_\textrm{M} - \varphi_{\textrm{M,ref}}\}
}{2 \pi}\right\}.
\label{eq:n}
\end{equation}
{\color{black}where $\varphi_{\textrm{M,ref}}$ the wrapped phase of MBI on the reference point.}
Whether a MBI can be successfully unwrapped depends on the uncertainty of the estimated phase ambiguity, denoted as the successful unwrapping (SU) criteria, which is defined as follows
\begin{equation}
\mid \hat{n}_\textrm{M} - {n}_\textrm{M,true} \mid < \frac{1}{2},
 \label{eq:su}
\end{equation}

If $\hat{n}_{\textrm{M},0}$ satisfies the SU criteria, {the phase residual of the MBI can be calculated as
\begin{equation}
\varphi_\textrm{res,M} = \textrm{wrap}\left\{\varphi_\textrm{M} - 2 \pi \hat{n}_{M,0} \right\}.
\label{eq:ph_res_M}
\end{equation} 

Using the same reference point during the spatial PU of the MBI, the phase residual of the MBI can be unwrapped. The final unwrapped phase of the MBI can be written as follows
\begin{equation}
\begin{aligned}
 \phi_\textrm{M} & = \phi_\textrm{res,M} + 2\pi \cdot\textrm{round} \left\{\frac{\frac{{B_{\perp,\textrm{M}}}}{{B_{\perp,\textrm{S}}}} \phi_{\textrm{S}} - \textrm{wrap}\{\varphi_\textrm{M} - \varphi_\textrm{M,ref} \}}{2\pi}\right\}.
\label{eq:spu2}
\end{aligned}
\end{equation}
}

The same approach will be conducted on LBI. If the phase ambiguity for LBI satisfies the SU criteria, the final unwrapped phase {\color{black}of the LBI} can be calculated as follows
\begin{equation}
\begin{aligned}
 \phi_\textrm{L} = \phi_\textrm{res,L} + 2\pi \cdot \textrm{round} \left\{\frac{{ \frac{{B_{\perp,\textrm{L}}}}{{B_{\perp,\textrm{M}}}} \phi_{\textrm{M}} } - \textrm{wrap}\{\varphi_\textrm{L} - \varphi_{\textrm{L,ref}}\}
}{2 \pi}\right\}.
\end{aligned}
\end{equation}
With the unwrapped phase $\phi_\textrm{L}$, the final height can be estimated using the follows~\cite{hanssen01}
\begin{equation}
\color{black}
\hat{h}_\textrm{L} = \frac{\lambda}{4 \pi} \frac{R \sin \theta}{B_{\perp,\textrm{L}}}
\phi_{L}. 
\label{eq:ph_h}
\end{equation}


\subsection{Optimal Baseline Design}
Baseline length has significantly impact on the result of 3D PU with very sparse acquisition. 
According to Eq.~(\ref{eq:spu2}), performance of the asymptotic 3D PU depends on two key indicators, i.e., baseline length and phase variance of the MB interferograms. 
Given the specified coherence, it is possible to get the optimal baseline length using the SU criteria. However, the SU criteria described in Eq.~(\ref{eq:su}) can not be directly used in the practical process due to the unknown value of the phase ambiguity. 
Nevertheless, it shows that the phase ambiguity can be estimated unambiguously if the phase bias induced by the noise is smaller than $\pi$. With the approximation that the interferometric phase without an ambiguity follows a Gaussian distribution, the expected phase variance should be $(\pi/u_\alpha)^2$ with a level of significance $\alpha$. 

Supposing that the baseline lengths of four interferograms obtained by the {\color{black}TDA-InSAR} are $B_{\perp,1}, B_{\perp,2}, B_{\perp,3}, B_{\perp,4}$ in an ascending order, 
the SU criteria can be rewritten as follow
\begin{equation}
\textrm{max}_{i = 2:4} \left\{{\frac{B_{\perp,{i}}^2}{B_{\perp,{i-1}}^2}\sigma^2_{\phi_{i-1}}+\sigma^2_{\phi_i}} \right\}<\left( \frac{\pi}{u_\alpha} \right)^2.
\label{eq:su_4}
\end{equation}

The SU criteria in Eq.~(\ref{eq:su_4}) can be used to optimize the parameters of the TDA-InSAR. 
In addition to the baseline length, the input parameters of this system includes 
coherence ($\gamma$), expected relative height precision ($\sigma_h^0$) and the maximal height difference ($\Delta h_\textrm{max}$). Then the performance of the MB interferograms can be evaluated using three decision variables, i.e., success rate of phase unwrapping (SR), relative height precision ($\sigma_h$) and height ambiguity ($h_\textrm{amb}$). 
Success rate of phase unwrapping is the foundation of the 3D reconstruction, which should be as high as possible.

If all interferograms can be correctly unwrapped, the final relative height precision only depends on the longest baseline length. It is necessary to guarantee the longest baseline length. 
Additionally, larger height ambiguity will increase the flexibility of the system, which enables larger height difference within the experimental area. With the same success rate of phase unwrapping, we expect as high height ambiguity as possible. 

From Eq.~(\ref{eq:su_4}), the baseline length of SBI has a low bound with a given LBI. On the satisfaction of the PC assumption, the baseline length of SBI should be as long as possible to achieve a good phase unwrapping performance. However, longer baseline length of SBI would decrease the height ambiguity, reducing the flexibility of the system. Thus, a trade-off between the height ambiguity and the success rate of phase unwrapping should be made. 
Along this direction, the optimal baseline design for the TDA-InSAR is defined as follows

\begin{equation}
\begin{aligned}
& \textrm{arg max} \{SR, h_\textrm{amb}\}~\textrm{and}~\textrm{arg min} \{ \sigma_h\} \\
 & \textrm{s.t.}~ SR = u^{-1} \left\{ \mathop{\textrm{max}}\limits_{i=2:4} \left( \pi /\sqrt{\left(\frac{B_i}{B_{i-1}}\right)^2 \sigma_{\phi_{i-1}}^2 + \sigma_{\phi_{i}}^2 }\right) \right\} > 1 - \alpha \\
 & h_\textrm{amb}= \frac{\lambda R \textrm{sin}\theta }{2 B_1}; 
 \sigma_\textrm{h}= \frac{\lambda R \textrm{sin}\theta }{4 \pi B_4} \sigma_{\varphi_\textrm{L}}; \sigma^2_\varphi = \frac{1 -\gamma^2}{2 \gamma^2} \\
& B_1 = \textrm{min} \left\{ B_{\perp,1}, B_{\perp,2}-i B_{\perp,1}\vert_{i\in \mathbb{Z}} ,\dots 
, B_{\perp,4}-i B_{\perp,3}\vert_{i\in \mathbb{Z}}
\right\} \\
& B_2 =\textrm{max} \left\{ B_{\perp,1}, \textrm{min} \left\{ B_{\perp,2}, B_{\perp,2}-i B_{\perp,1}\vert_{i\in \mathbb{Z}} ,\dots
\right\} \right\} \\
& B_3 =\textrm{max} \left\{ B_{\perp,2}, \textrm{min} \left\{ B_{\perp,3}, B_{\perp,2}-i B_{\perp,1}\vert_{i\in \mathbb{Z}} ,\dots
\right\} \right\} \\
& B_4 = B_{\perp,4}; 
B_1 < \frac{\lambda R \textrm{sin}\theta}{2 \Delta h_\textrm{max}}, \sigma_h < \sigma_h^0
\label{eq:ob}
\end{aligned}
\end{equation}
where $B_1\sim B_4$ denote the effective baseline length. In this optimization, the relationship between the coherence and the phase variance is approximated for point scatterer with a coherence larger than 0.9~\cite{hanssen01a}. Note that the model is also simplified by assuming that the coherence corresponding to LBI, MBI and SBI are the same, which is suitable for point scatterers. For distributed scatterers, other factors, such as spatial and volume decorrelation should be considered during the optimization, which is not investigated in this paper.

\subsection{Height Inversion with the Error Compensation}
The {\color{black}dual-satellite} interferogram 
enables a long baseline but suffers from significant orbit error, which will bias the final height estimation. Fortunately, with the help of the accurate initialization obtained by the dual-antenna interferogram, the {\color{black}dual-satellite} interferogram can be successfully unwrapped since the phase change induced by the orbit error will not destroy the PC assumption. Therefore, the orbit error can be well compensated during the height inversion. 
 


\begin{figure}[htbp]
     \centering
     \includegraphics[width=0.35\textwidth]{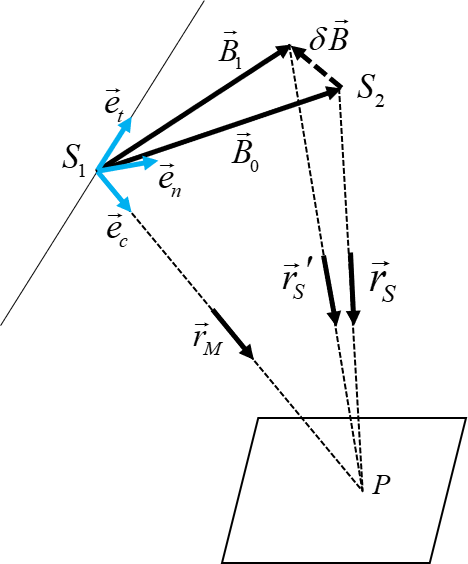}
     \caption{ Geometry of the InSAR baseline error.}
     \label{fig:Be}    
\end{figure} 

{\color{black}Here the nonlinear baseline model is used to model the orbit error. }
Generally, the orbit errors in both master and slave images can be expressed as the baseline error. 
Fig.~\ref{fig:Be} shows the geometry of the baseline error. The additional phase change induced by the orbit error can be defined as 
\begin{equation}
\phi_\textrm{orb}=\frac{4 \pi}{\lambda} \vert {\vec{r}_S^{\prime}} - {\vec{r}_S} \vert,
\end{equation}
where ${\vec{r}_S^{\prime}}$ and ${\vec{r}_S}$ denote the actual and error-free slave distance vector along the line of sight. Note that ${\vec{r}_S^{\prime}}$ also includes the orbit error in the master distance vector $\vec{r}_{M}$. 
In the practical process, the baseline vector $\vec{B}$ in a TCN (track, cross, normal) representation is often used, denoted as $(B_t, B_c, B_n)$. Then the orbit error as a function of baseline vector can be written as

\begin{equation}
\vert {\vec{r}_S^{\prime}} - {\vec{r}_S} \vert = \frac{\partial{\vec{r}_S}}{\partial{\vec{B}}} = \frac{\partial{\vec{r}_S}}{\partial{{B}_t}} \vec{e}_t
+ \frac{\partial{\vec{r}_S}}{\partial{{B}_c}} \vec{e}_c + \frac{\partial{\vec{r}_S}}{\partial{{B}_n}} \vec{e}_n,
\end{equation}
where the partial deviation of the baseline component can be expressed as follows
\begin{equation}
\frac{\partial{\vec{r}_S}}{\partial{{B}_t}} =
\frac{ \vert B_t \vert - \vec{r}_M \cdot \vec{e}_t }
{
\sqrt{
{\vert \vec{r}_M \vert}^2 + {\vert \vec{B} \vert}^2 - 2 \cdot \vec{r}_M \cdot \vec{B}
}
}.
\end{equation}

The relationship between the orbit error phase and baseline error can be written as
\begin{equation}
E\{\phi_\textrm{orb}\} = \frac{4 \pi} {\lambda} \left(
\frac{\partial{\vec{r}_S}}{\partial{{B}_t}} \delta B_t 
+ \frac{\partial{\vec{r}_S}}{\partial{{B}_c}} \delta B_c 
+ \frac{\partial{\vec{r}_S}}{\partial{{B}_n}} \delta B_n \right)
\end{equation}
Since the orbit error may change temporally during the radar image focusing, the first-order term $\partial{\dot{\vec{B}}}$ should be added to model the fringe residuals. Additionally, for small squint angles in spaceborne SAR, the interferometric phase is not sensitive to the orbit errors along the track direction. So the orbit error components along the track direction are neglected. The orbit error phase can be expressed as follows
\begin{equation}
\begin{aligned}
E\{\phi_\textrm{orb}\} = & \frac{4 \pi} {\lambda} \left (
\frac{\partial{\vec{r}_S}}{\partial{{B}_c}} (\delta B_c + \delta \dot{B_c} t ) 
+ \frac{\partial{\vec{r}_S}}{\partial{{B}_n}} (\delta B_n + \delta \dot{B_n} t) \right )
\end{aligned}
\label{eq:orb_model}
\end{equation}

If the TDA-InSAR works in a bi-static mode, the orbit errors of both MBI and LBI have the same spatial variation but different value. Additionally, the atmospheric delay can be totally neglected due to the strict synchronization.
With the unwrapped phases $\phi_\textrm{M}$ and $\phi_\textrm{L}$, the parameters related to the orbit error can be estimated jointly with the height. 
Supposing that there are $m$ coherent scatterers, the mathematical model can be defined as follows

\begin{equation}
\begin{split}
\begin{bmatrix}
{\phi}_{\textrm{M},1} \\ 
  \vdots\\
{\phi}_{\textrm{M},m} \\
{\phi}_{\textrm{L},1} \\ 
  \vdots\\
{\phi}_{\textrm{L},m} \\
\end{bmatrix}
& = \frac{4\pi}{\lambda}
\left [
\begin{matrix}
  {\textbf{B}_\textrm{orb}^\textrm{M}} 
  & \begin{matrix}
  \frac{R_1 \sin \theta_1}{B_{\perp,1}^\textrm{M}}  & \hdots & 0 \\
  \vdots  & \ddots & \vdots\\
 0 &  \hdots &  \frac{R_n \sin \theta_n}{B_{\perp,n}^\textrm{M}} \\
  \end{matrix}\\
  {\textbf{B}_\textrm{orb}^\textrm{L}} 
  & \begin{matrix}
  \frac{R_1 \sin \theta_1}{B_{\perp,1}^\textrm{L}} &  \hdots & 0 \\
  \vdots  & \ddots & \vdots\\
 0  & \hdots &  \frac{R_m \sin \theta_m}{B_{\perp,m}^\textrm{L}} \\
  \end{matrix}
\end{matrix}
 \right ]
\begin{bmatrix}
\delta B_c \\
\delta \dot{B_c} \\
\delta B_n \\
\delta \dot{B_n} \\
  h_1 \\
  h_2 \\
  \vdots \\
  h_m \\
  \end{bmatrix}
\end{split}
\label{eq:model}
\end{equation} 

where 
\begin{equation}
\textbf{B}_\textrm{orb}=
\begin{bmatrix}
    \frac{\partial{\vec{r}_S}}{\partial{{B}_{c,1}}} & \frac{\partial{\vec{r}_S}}{\partial{{B}_{c,1}}} t_1  & \frac{\partial{\vec{r}_S}}{\partial{{B}_{n,1}}} t_1  & \frac{\partial{\vec{r}_S}}{\partial{{B}_{n,1}}} t_1  \\
    \vdots & \vdots & \vdots & \vdots \\
    \frac{\partial{\vec{r}_S}}{\partial{{B}_{c,m}}} & \frac{\partial{\vec{r}_S}}{\partial{{B}_{c,m}}} t_m  & \frac{\partial{\vec{r}_S}}{\partial{{B}_{n,m}}} t_m  & \frac{\partial{\vec{r}_S}}{\partial{{B}_{n,m}}} t_m \\
  \end{bmatrix}
\end{equation} 

The expression in Eq.~(\ref{eq:model}) represents a linear system of $2m$
equations in $m+4$ unknowns. Thus a unique solution can be obtained for such determined problem. 

If the TDA-InSAR works in a mono-static mode, the orbit error of the mono-static dual-satellite interferogram will be
twice than that of the bi-static one. If the MBI is a bi-static interferogram and LBI is a mono-static interferogram, the coefficient $\textbf{B}_\textrm{orb}^{\textrm{M}}$ equals to $\textbf{B}_\textrm{orb}^{\textrm{L}}/2$.

Research in~\cite{Hu2022DALES} shows that the atmospheric delay, especially the tropospheric delay will decorrelate to a half-time value within 3 minutes. Thus, if the TDA-InSAR works in a mono-static mode, both orbit error and atmospheric delay should be considered during the height estimation.
The atmospheric delay should be calculated per pixel due to the high spatial correlation{\color{black}.}
The mathematical model with respect to the atmospheric delay correction is defined as follows
\begin{equation}
\begin{split}
\begin{bmatrix}
{\phi}_{\textrm{S},1} \\ 
  \vdots\\
{\phi}_{\textrm{S},m} \\
{\phi}_{\textrm{M},1} \\ 
  \vdots\\
{\phi}_{\textrm{M},m} \\
{\phi}_{\textrm{L},1} \\ 
  \vdots\\
{\phi}_{\textrm{L},m} \\
\end{bmatrix}
& = \frac{4\pi}{\lambda}
\left [
\begin{matrix}
  \bf{0} 
  & \begin{matrix}
  \frac{R_1 \sin \theta_1}{B_{\perp,1}^\textrm{S}}  & \hdots & 0 \\
  \vdots  & \ddots & \vdots\\
 0 &  \hdots &  \frac{R_m \sin \theta_m}{B_{\perp,m}^\textrm{S}} \\
  \end{matrix}
  & \bf{0}
  \\
  {\textbf{B}_\textrm{orb}^{\textrm{M}}} 
  & \begin{matrix}
  \frac{R_1 \sin \theta_1}{B_{\perp,1}^\textrm{M}}  & \hdots & 0 \\
  \vdots  & \ddots & \vdots\\
 0 &  \hdots &  \frac{R_m \sin \theta_m}{B_{\perp,m}^\textrm{M}} \\
  \end{matrix} &
  \textbf{I}_m 
  \\
 {\textbf{B}_\textrm{orb}^{\textrm{L}}} 
  & \begin{matrix} 
  \frac{R_1 \sin \theta_1}{B_{\perp,1}^\textrm{L}} &  \hdots & 0 \\
  \vdots  & \ddots & \vdots\\
 0  & \hdots &  \frac{R_m \sin \theta_m}{B_{\perp,m}^\textrm{L}} \\
  \end{matrix} &
    \textbf{I}_m 
  \\
\end{matrix}
 \right ]
\begin{bmatrix}
\delta B_c \\
\delta \dot{B_c} \\
\delta B_n \\
\delta \dot{B_n} \\
  h_1 \\
  \vdots \\
  h_m \\
  D_{1} \\
  \vdots \\
  D_{m} \\
  \end{bmatrix}
\end{split}
\label{eq:model}
\end{equation} 
{\color{black}where $\textbf{I}_m$ is a $m\times m$ identity matrix.} $D_1,\cdots,D_m$ denotes atmospheric delay of the $m$ scatterers.

\section{Investigation of different TDA-InSAR baseline configurations}
\label{sec:3}
\begin{figure*}[htbp]
\centering
     \begin{subfigure}[b]{.24\textwidth}
  \includegraphics[width=\textwidth]{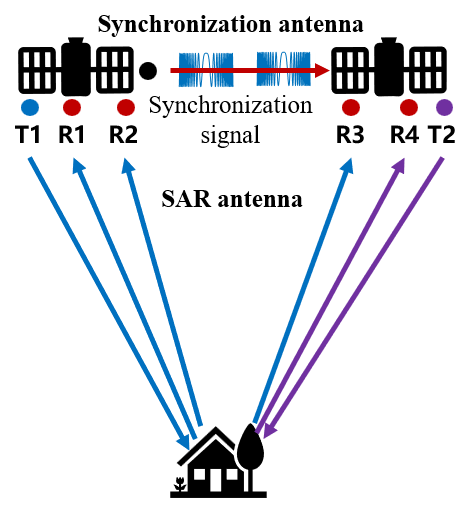}
  \caption{baseline configuration 1}
\end{subfigure}
\begin{subfigure}[b]{.24\textwidth}
  \includegraphics[width=\textwidth]{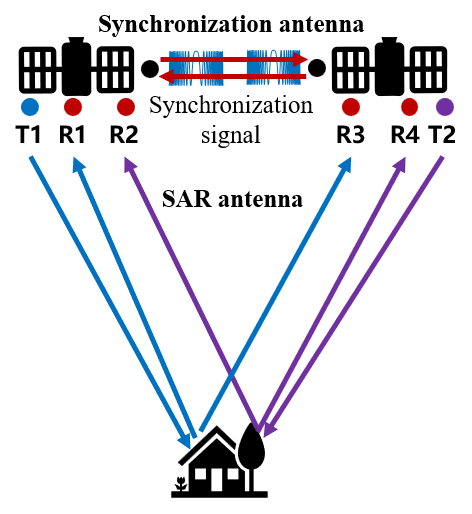}
  \caption{baseline configuration 2}
\end{subfigure} 
     \begin{subfigure}[b]{.24\textwidth}
  \includegraphics[width=\textwidth]{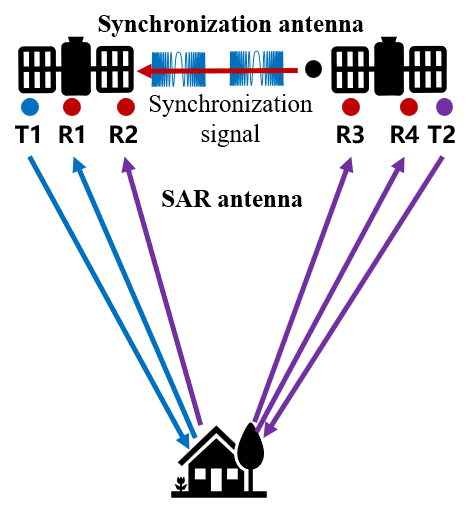}
  \caption{baseline configuration 3}
\end{subfigure}
\begin{subfigure}[b]{.24\textwidth}
  \includegraphics[width=\textwidth]{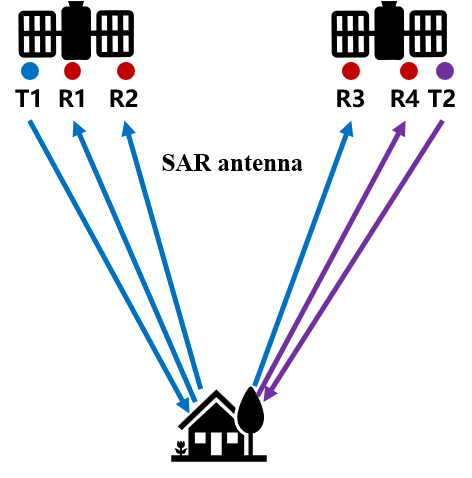}
  \caption{baseline configuration 4}
\end{subfigure} 
     \caption{\color{black} TDA-InSAR with different baseline configurations. (a) (b) and (c) are bistatic modes. (d) is mono-static mode. }
       \label{fig:mode_geometry}   
\end{figure*}

{\color{black}\subsection{Definition of different baseline configuration}}
Considering different types of data transmission and reception, the proposed TDA-InSAR could work in either bistatic or mono-static modes with varying baseline configurations. 
Supposing there are two transmission radars, denoted as T1 and T2 and four reception radars, denoted as R1$\sim$R4. 
In order to get an interferogram with the longest baseline, two acquisitions obtained by T1-R1 and T2-R4 must be selected. 
Then there are four types of baseline configurations for the other acquisitions, as shown in 
{\color{black}Fig.~\ref{fig:mode_geometry}. } 
Note that the baseline configurations 1, 2 and 3 are bistatic modes while the fourth baseline configuration is mono-static mode. 

{\color{black}
If the bi-static mode is used, signal synchronization between the two satellites is required, which may be achieved by using beam, time and phase synchronizations~\cite{2008Bistatic}. TanDEM-X achieves the beam synchronization by ensuring both antennas can illuminate the same area or employing zeros Doppler steering~\cite{Krieger2007}. Time synchronization gives the precise time when a chirp signal is transmitted, which can be achieved using leap pulse repetition intervals method~\cite{Krieger2007} or GPS disciplined rubidium clock USOs for time calibration~\cite{Jin2020}. Moreover, phase synchronization must be implemented to guarantee the coherence between the transmitter and receiver. Additional antennas covering the full solid angle are used for a mutual exchange of phase synchronization signal between the two satellites. Details can be found in~\cite{Jin2020,Liang2020}. 
In the practical application, the baseline configurations 1 and 3 need to transmit the synchronous signal in a single direction while the baseline configuration 2 requires a bidirectional synchronization. 
}

Assuming that the antenna baselines of the two satellites are the same, the equivalent baselines of different baseline configurations can be obtained by setting the acquisition T1-R1 as the master image. The equivalent baselines of the baseline configuration 1 are as follows 
\begin{equation}
B_{\perp,1} = L_1 / 2;B_{\perp,2} = L_2 / 2;B_{\perp,3} = L_2 + L_1, 
\end{equation}
those of the baseline configuration 2 are 
\begin{equation}
B_{\perp,1} = L_2 / 2;B_{\perp,2} = L_2 / 2 + L_1;B_{\perp,3} = L_2 + L_1,
\end{equation}
those of the baseline configuration 3 are
\begin{equation}
B_{\perp,1} = L_1 +L_2/ 2 ;B_{\perp,2} = L_2 + L_1/ 2;B_{\perp,3} = L_2 + L_1,
\end{equation}
and those of the baseline configuration 4 are 
\begin{equation}
B_{\perp,1} = L_1 / 2;B_{\perp,2} = L_2 +L_1/2;B_{\perp,3} = L_2 + L_1.
\label{eq:mode4}
\end{equation}

Given a specified coherence, the SU criteria gives an upper bound of the baseline ratio. 
In order to unwrap the interferogram with as long baseline as possible, another interferogram with a medium baseline is required. 
Since the baseline length $B_{\perp,2}$ is almost the same as $B_{\perp,3}$ in Eq.~(\ref{eq:mode4}), the performance of mono-static mode is theoretically poorer than that of the bistatic mode due to the lack of medium baseline interferogram. 


Considering the baseline configurations 2 and 3, the BLC approach should be applied to get a pseudo short baseline interferogram. But the combination integer of the baseline configuration 3 is larger than that of the baseline configuration 2, leading to a noisy initial height estimation. Theoretically, the baseline configuration 2 is better than the baseline configuration 3. 
Note that MB interferograms obtained by all baseline configurations have a large height ambiguity due to the ultra short baseline interferogram. 
Further simulation-based performance analysis of different baseline configurations will be conducted using the optimal baseline design in Eq.(\ref{eq:ob}).

\begin{table}
\caption{Main parameters for the simulation}
\begin{tabularx}{\linewidth}{lXX}
\toprule
Parameters & Values\\
\midrule
Frequency (GHz) & 9.6 \\
Resolution (range $\times$ azimuth, m) & 0.93 $\times$ 2.00 \\
Incident angle (degree) & 30 \\ 
Near range (m) &  608015 \\ 
Maximal antenna baseline (m) & 20 \\
Maximal satellite baseline (m) & 500 \\
Expected height precision (m) & 0.5 \\
\bottomrule
\end{tabularx}
\label{tab:1}
\end{table}

\begin{figure}[htbp]
     \centering
     \includegraphics[width=0.4\textwidth]{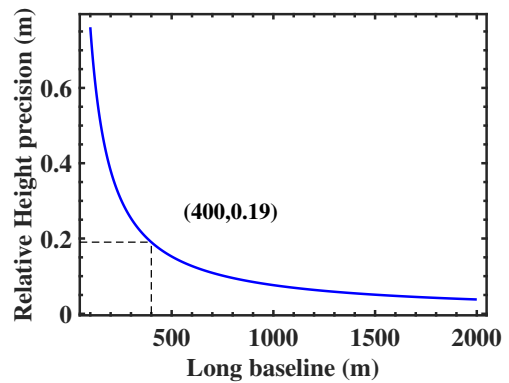}
     \caption{ Relative height precision as a function of long baseline length..}
     \label{fig:hstd}    
\end{figure}  

{\color{black}\subsection{Performance analysis}}

\begin{figure*}[htbp]
\centering
     \begin{subfigure}[b]{.24\textwidth}
  \includegraphics[width=\textwidth]{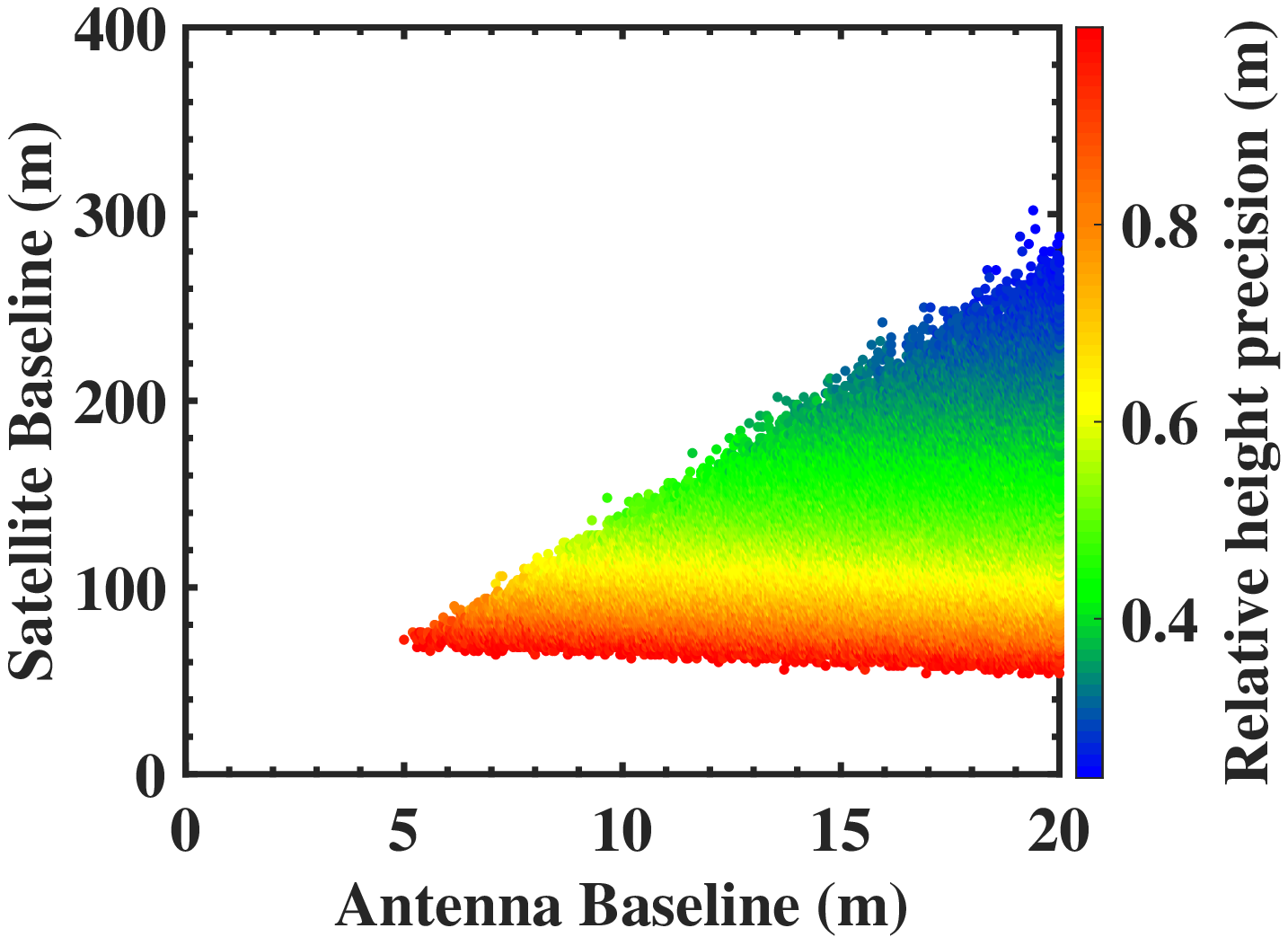}
  \caption{baseline configuration 1}
\end{subfigure}
\begin{subfigure}[b]{.24\textwidth}
  \includegraphics[width=\textwidth]{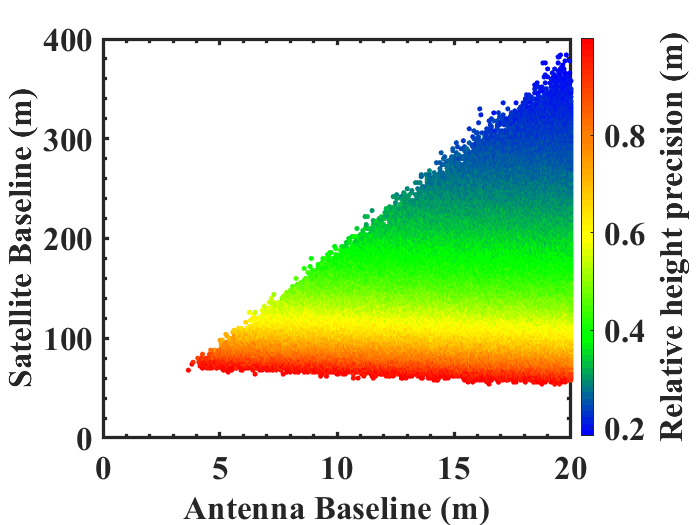}
  \caption{baseline configuration 2}
\end{subfigure}
\begin{subfigure}[b]{.24\textwidth}
  \includegraphics[width=\textwidth]{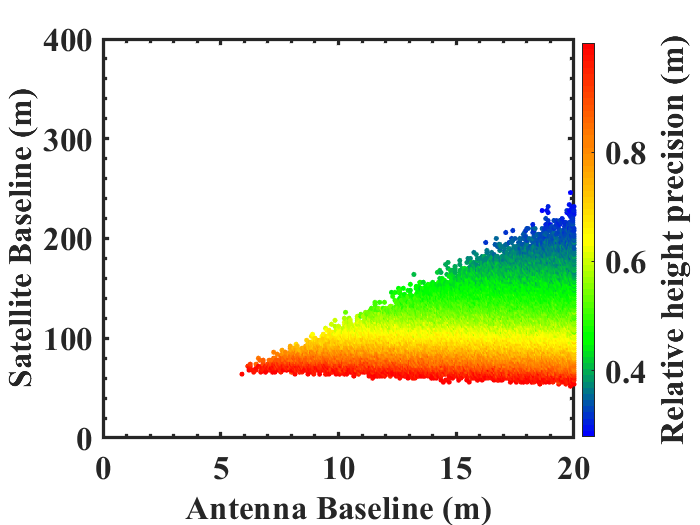}
  \caption{baseline configuration 3}
\end{subfigure}
\begin{subfigure}[b]{.24\textwidth}
  \includegraphics[width=\textwidth]{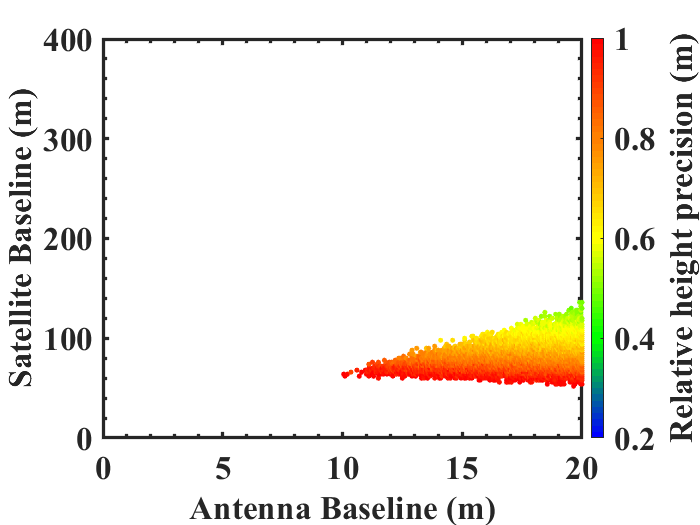}
  \caption{baseline configuration 4}
\end{subfigure}
\begin{subfigure}[b]{.24\textwidth}
  \includegraphics[width=\textwidth]{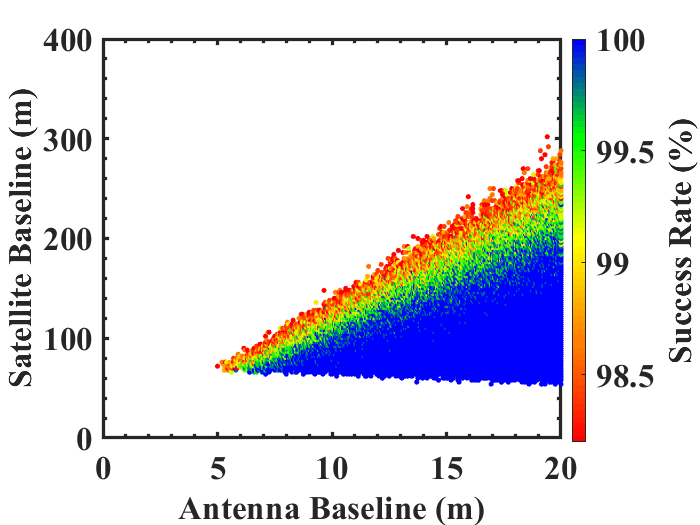}
  \caption{baseline configuration 1}
\end{subfigure} 
\begin{subfigure}[b]{.24\textwidth}
  \includegraphics[width=\textwidth]{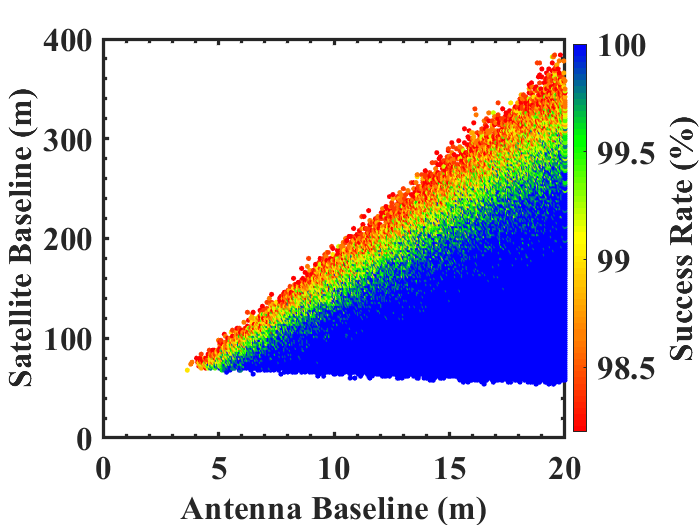}
  \caption{baseline configuration 2}
\end{subfigure}
\begin{subfigure}[b]{.24\textwidth}
  \includegraphics[width=\textwidth]{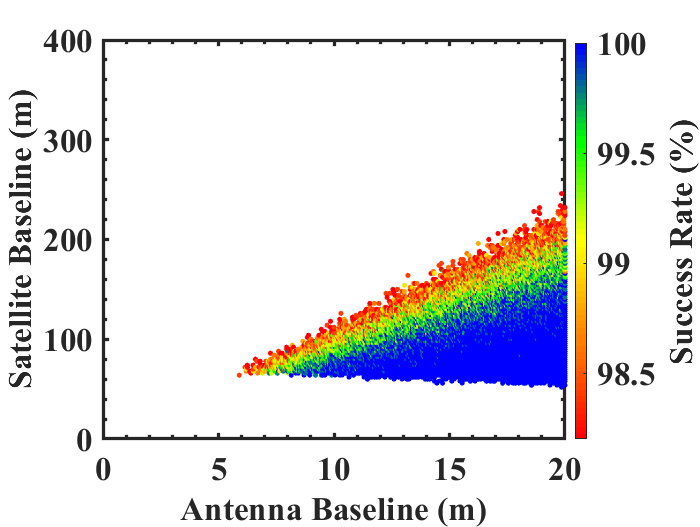}
  \caption{baseline configuration 3}
\end{subfigure}
\begin{subfigure}[b]{.24\textwidth}
  \includegraphics[width=\textwidth]{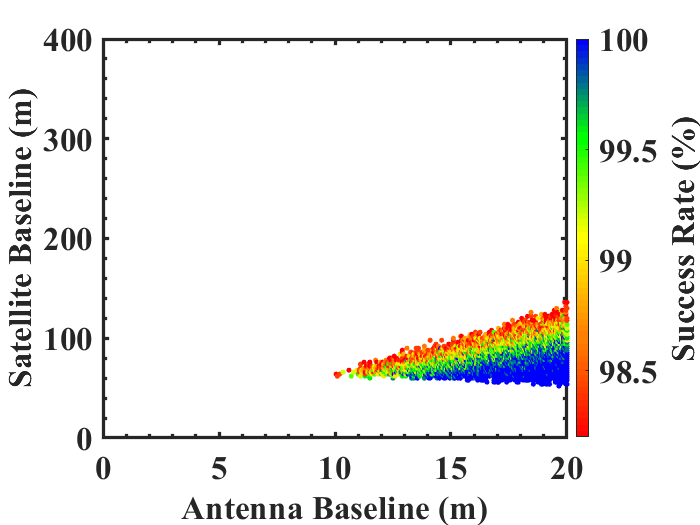}
  \caption{baseline configuration 4}
\end{subfigure}
     \caption{ Performance of 3D reconstruction using different baseline configurations. First row: relative height precision; Second row: success rate of phase unwrapping. }
       \label{fig:mode}   
\end{figure*}

The simulated parameters are listed in Table.~\ref{tab:1}. 
Setting the coherence to 0.99, the relationship between the long baseline length and the relative height precision is shown in Fig.~\ref{fig:hstd}. If all interferograms are unwrapped successfully,the relative height precision will not increase significantly after the baseline length reaches 400 m. 
Then a comparative study is conducted to evaluate the performance of 3D reconstruction using different baseline configurations. 
In the following simulation, the antenna baseline ranges from 0.5 to 20 m with a step of 0.1 m, the satellite baseline is from 10 to 400 m with a step of 2 m and the expected SR is set to 0.98. 
For each MB interferograms, the simulation is repeated 500 times and the corresponding relative height precision is obtained correspondingly. Fig.~\ref{fig:mode} denotes the relative height precision and success rate of phase unwrapping using different baseline configurations.  

\begin{figure*}[htbp]
\centering
\begin{subfigure}[b]{.32\textwidth}
  \includegraphics[width=\textwidth]{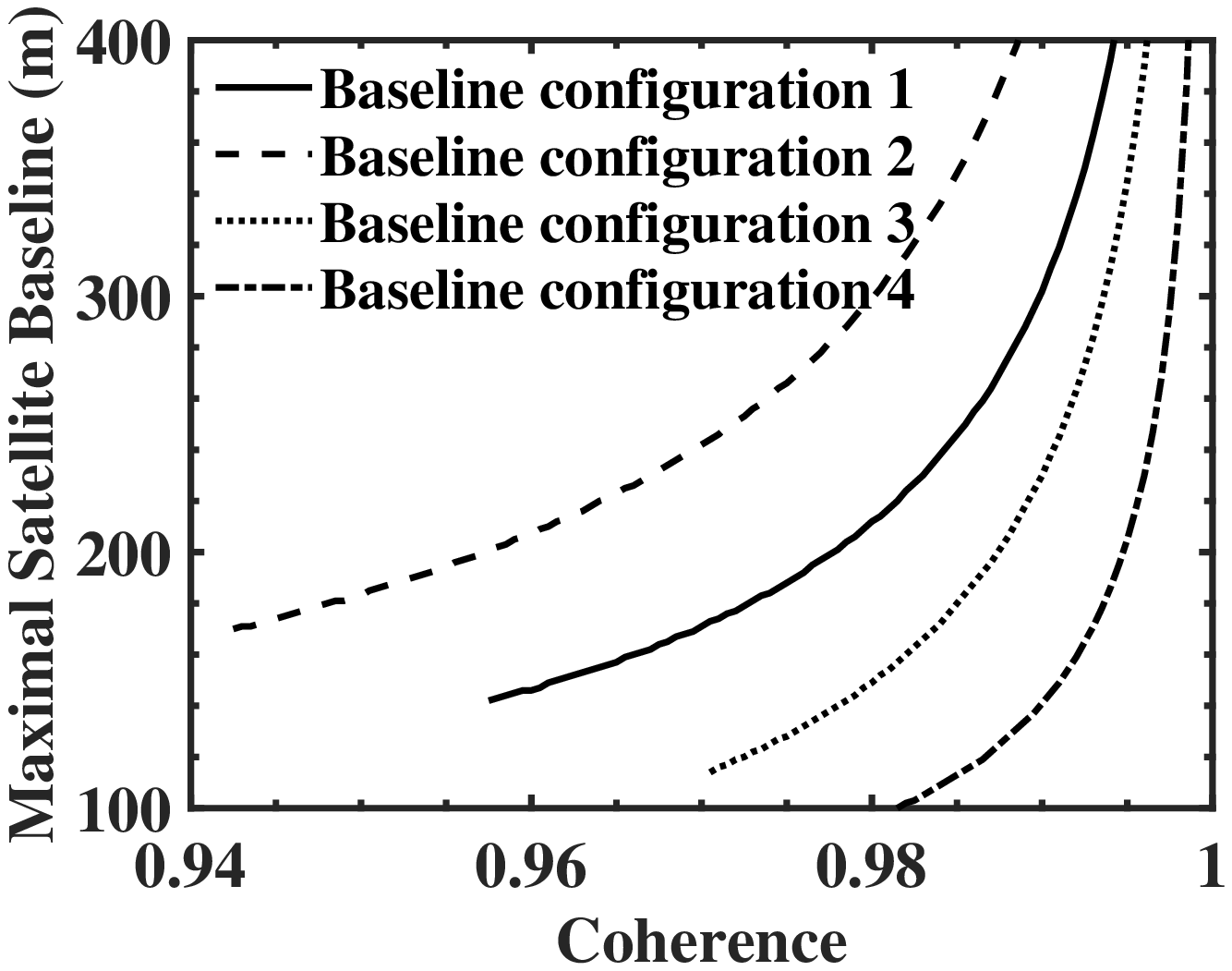}
  \caption{}
\end{subfigure}
\begin{subfigure}[b]{.32\textwidth}
  \includegraphics[width=\textwidth]{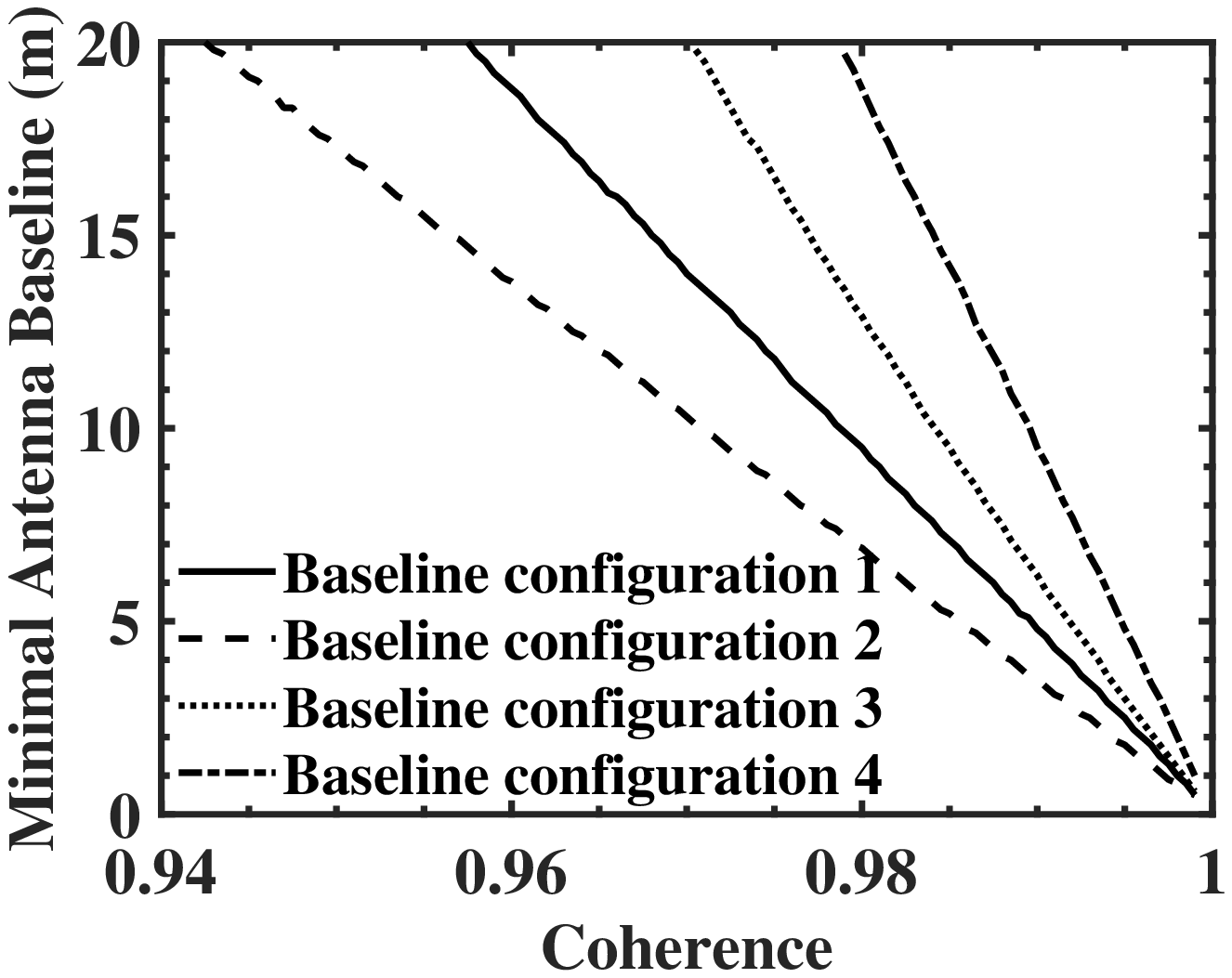}
  \caption{}
\end{subfigure}
\begin{subfigure}[b]{.32\textwidth}
  \includegraphics[width=\textwidth]{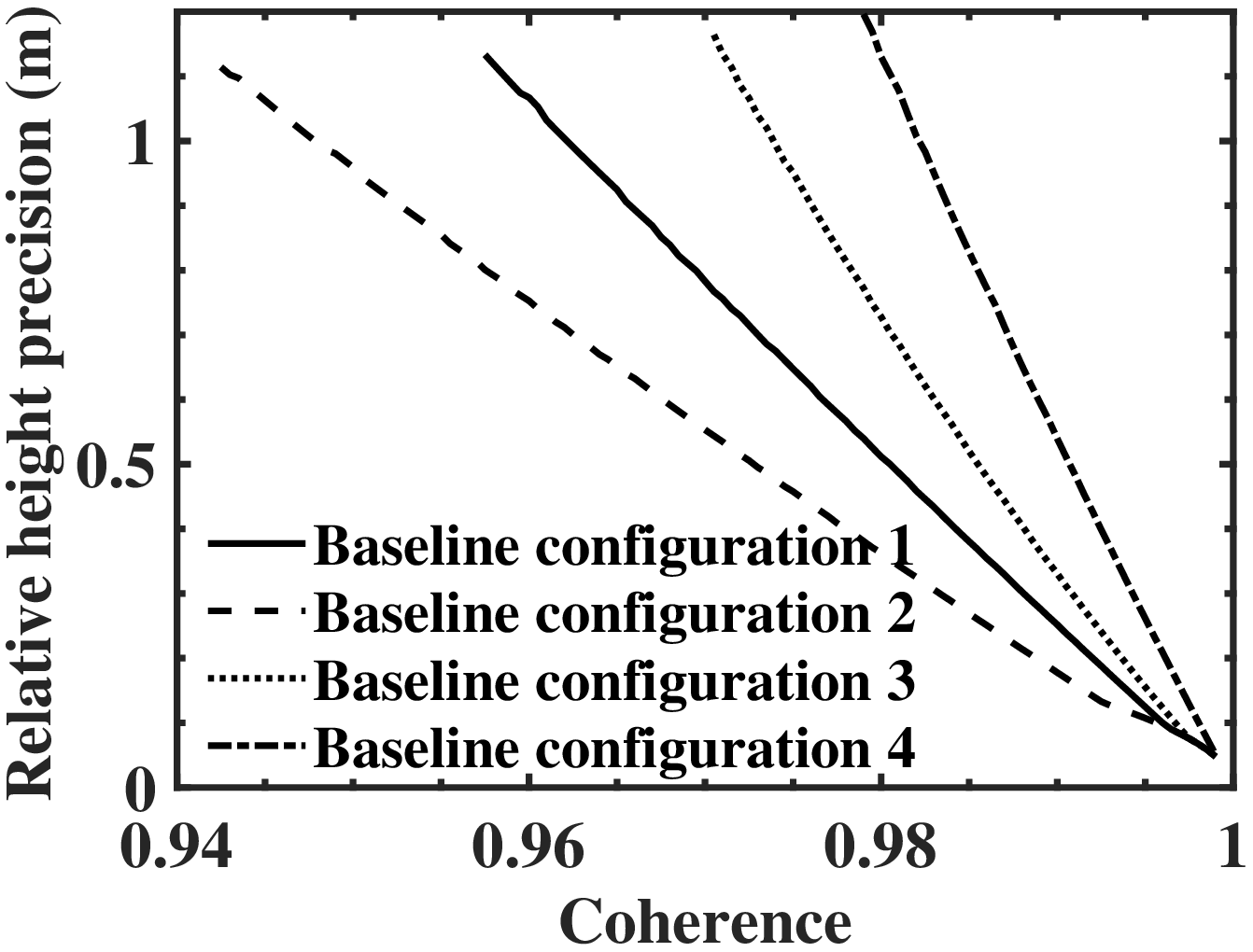}
  \caption{}
\end{subfigure} 
     \caption{ Performance of 3D reconstruction using different baseline configurations. }
       \label{fig:mode1}   
\end{figure*}

\begin{figure*}[htbp]
\centering
\begin{subfigure}[b]{.3\textwidth}
  \includegraphics[width=\textwidth]{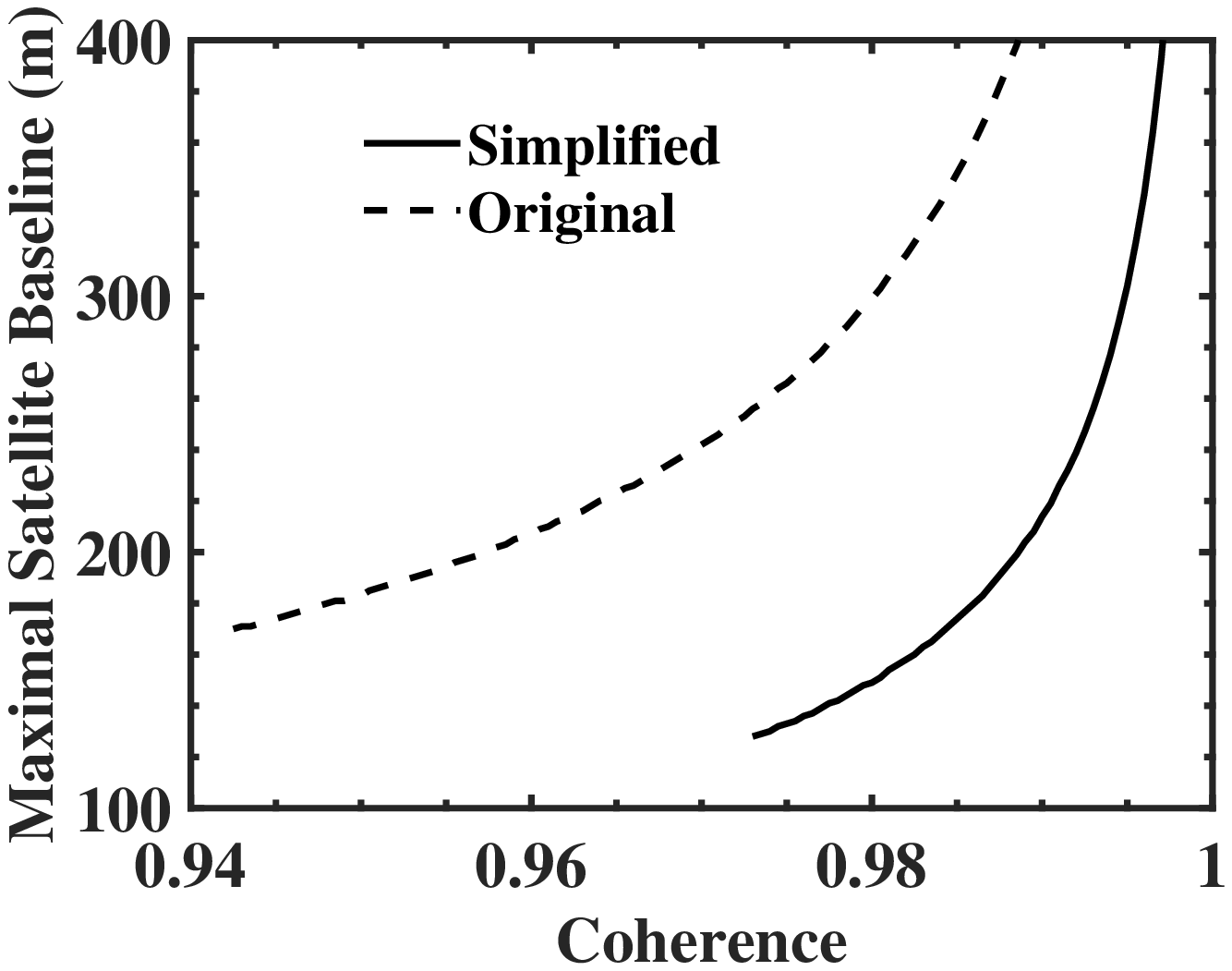}
  \caption{}
\end{subfigure}
\begin{subfigure}[b]{.3\textwidth}
  \includegraphics[width=\textwidth]{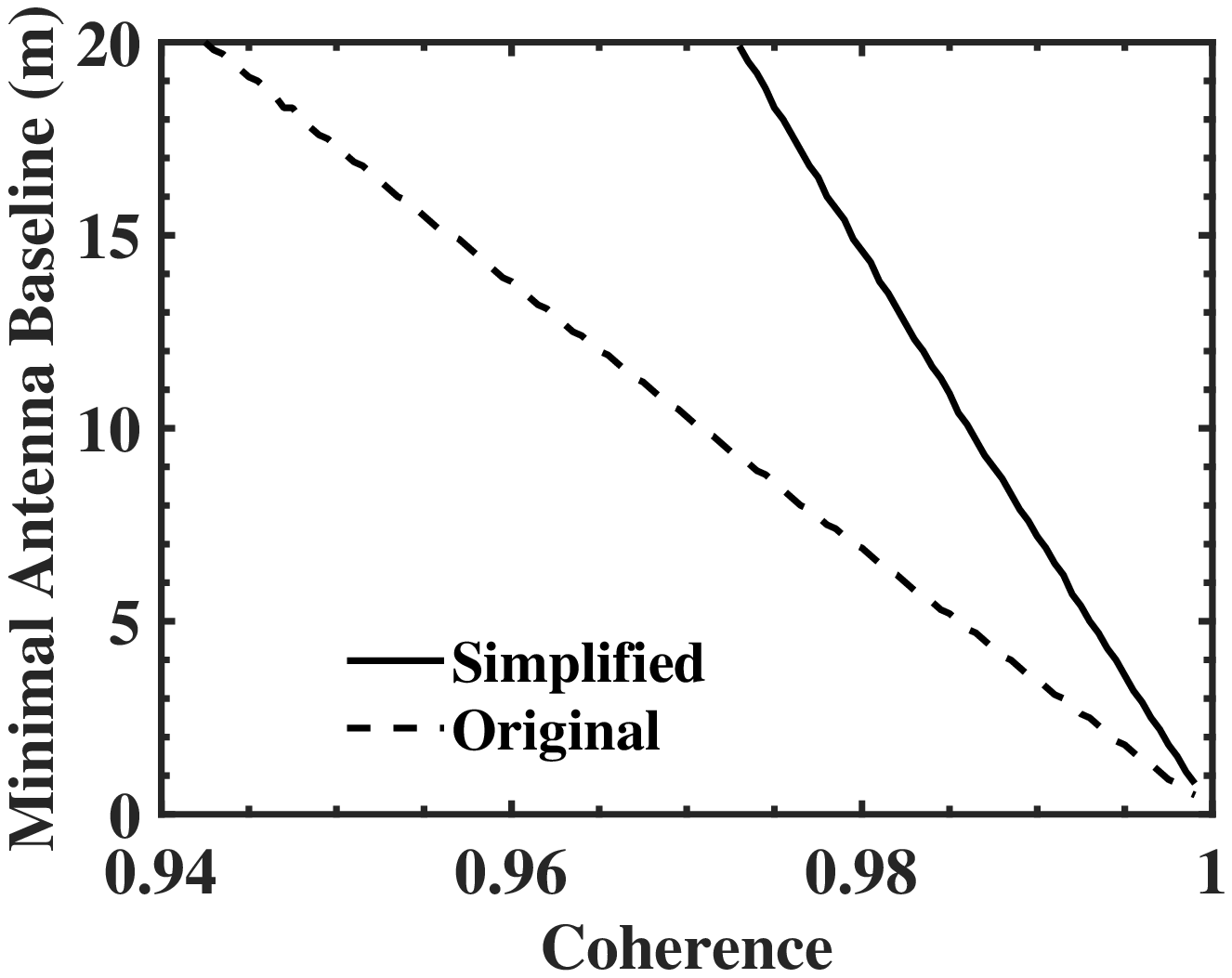}
  \caption{}
\end{subfigure}
\begin{subfigure}[b]{.3\textwidth}
  \includegraphics[width=\textwidth]{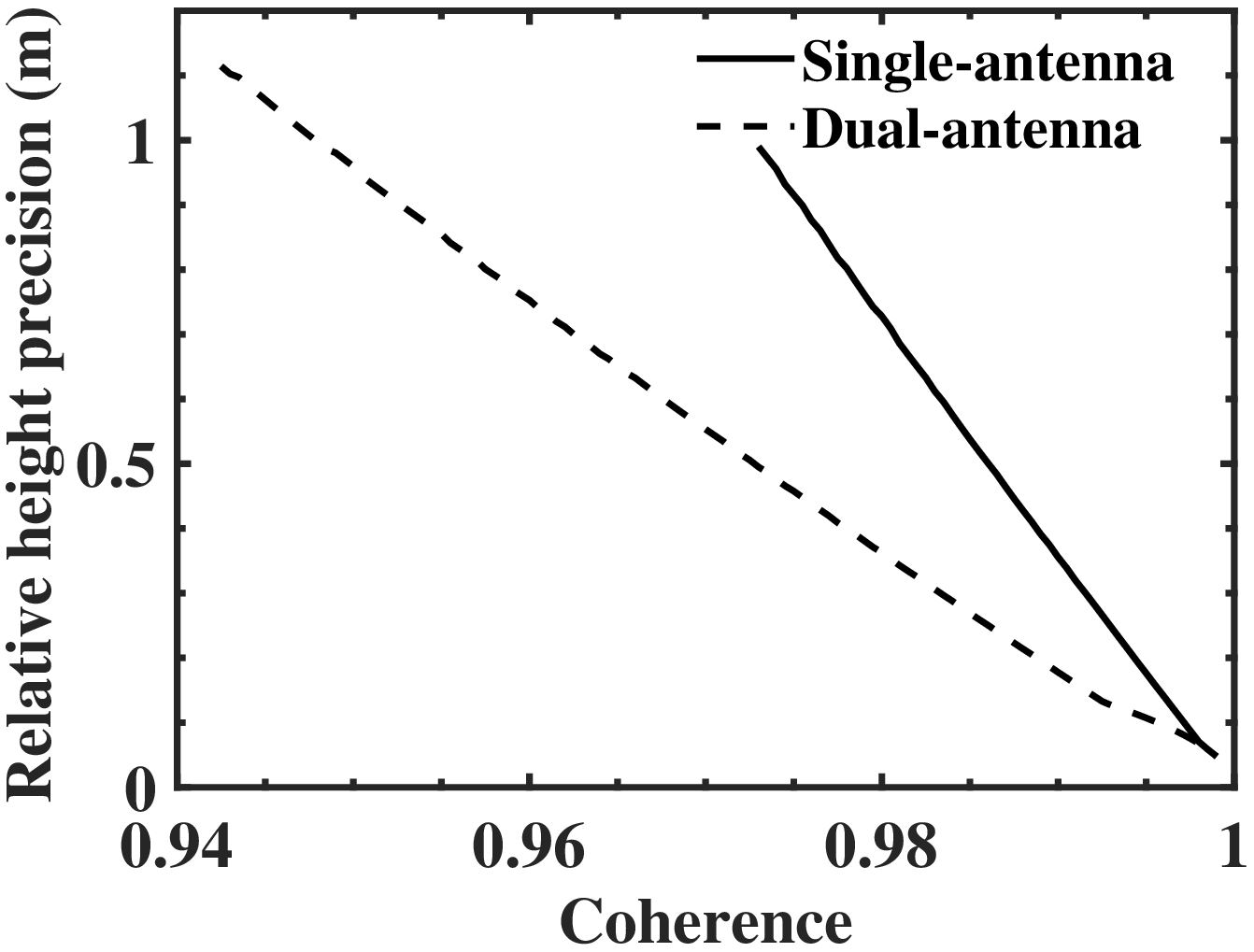}
  \caption{}
\end{subfigure} 
     \caption{ Performance comparison between simplified and original TDA-InSAR. }
       \label{fig:mode2}   
\end{figure*}

Fig.~\ref{fig:mode} (d) shows that the longest baseline is 100 m for the MB interferograms obtained by the mono-static mode. 
On the contrary, the bi-static measurements between different satellites provide a transition interferogram with a medium baseline, which enables the successful phase unwrapping of LBI with a baseline length of 200 m, as shown in Figs.~\ref{fig:mode} (a) (b) and (c). Additionally, MB interferograms obtained by the baseline configuration 2 show the most flexible baseline design while that by the baseline configuration 4 show a very strict baseline design. For example, assuming that the antenna baseline is 10 m, the possible satellite baseline of the baseline configuration 2 ranges from 50 to 200 m with the expected relative height precision of 1 m while that of the baseline configuration 4 only succeeds with a satellite baseline of 60 m. 
Due to the instability of the platform, the baseline of {\color{black}dual-satellite} interferogram may be inaccurate. Larger range of available baseline design is able to increase successful rate of phase unwrapping in the practical application. 
Furthermore, given the same coherence, the minimal antenna baseline of the baseline configuration 2 is 3.8 m while that of the baseline configuration 4 is 10 m. Shorter antenna would decrease the complexity of the system design.

Since the success rate of phase unwrapping varies with the coherence, the maximal satellite baseline, minimal antenna baseline and relative height precision as a function of coherence are obtained, separately, as shown in Fig.~\ref{fig:mode1}. 
According to Figs.~\ref{fig:mode1} (a) and (c), the maximal satellite baseline of the baseline configuration 2 is longer than other baseline configurations, leading to a better relative height precision. Moreover, a shorter antenna length can be used using the baseline configuration 2 with the same performance. 



Furthermore, if this system only contains a dual-antenna and a single-antenna satellite, i.e., two interferograms in a single flight, it is also possible to achieve a 3D reconstruction using the simplified system. If the baseline configuration 2 is used, the three channel images are obtained by T1-R1, T2-R2, T2-R4. 
A performance comparison between the original TDA-InSAR and the simplified one is conducted,  
as shown in Fig.~\ref{fig:mode2}. 
Fig.~\ref{fig:mode2} shows that the performance of the simplified TDA-InSAR is only half than that of the original one.
For example, if the coherence is 0.98, the maximal satellite baseline of the TDA-InSAR is 300 m while that of the simplified one is 150 m. Similar conclusions can be found in the analysis of both minimal antenna baseline and relative height precision. Thus, the additional antenna significantly reduce the expected coherence, leading to less difficulties in the practical implementation. 

\begin{figure}[htbp]
     \centering
     \includegraphics[width=0.45\textwidth]{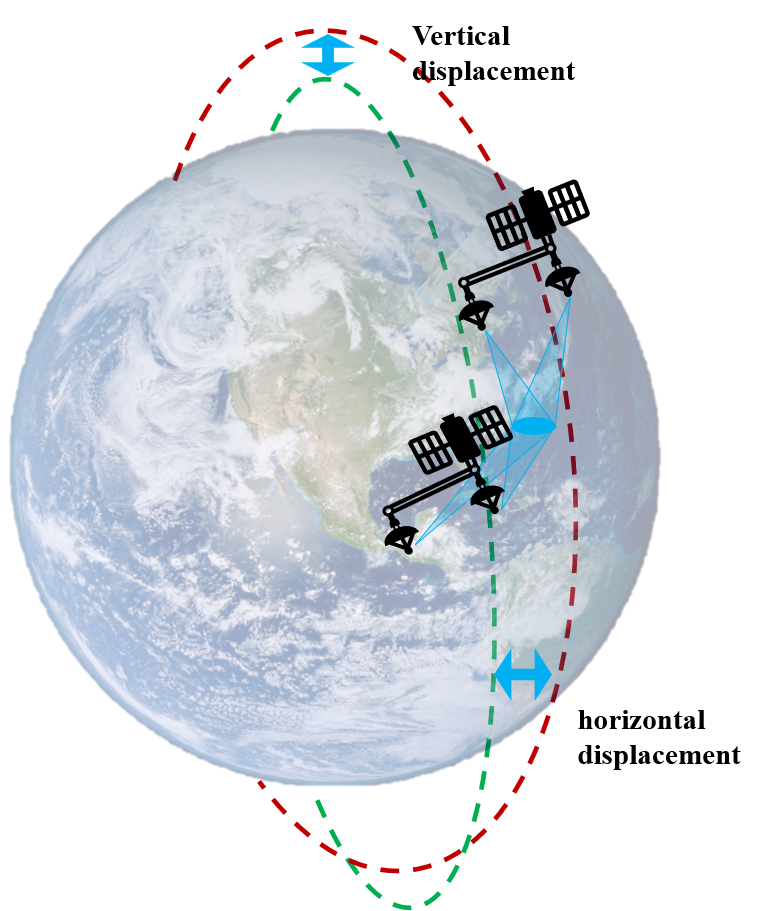}
     \caption{ Sketch of the Helix satellite formation.}
     \label{fig:helix}    
\end{figure}  

\begin{figure}[htbp]
\centering
     \begin{subfigure}[b]{.48\textwidth}
  \includegraphics[width=\textwidth]{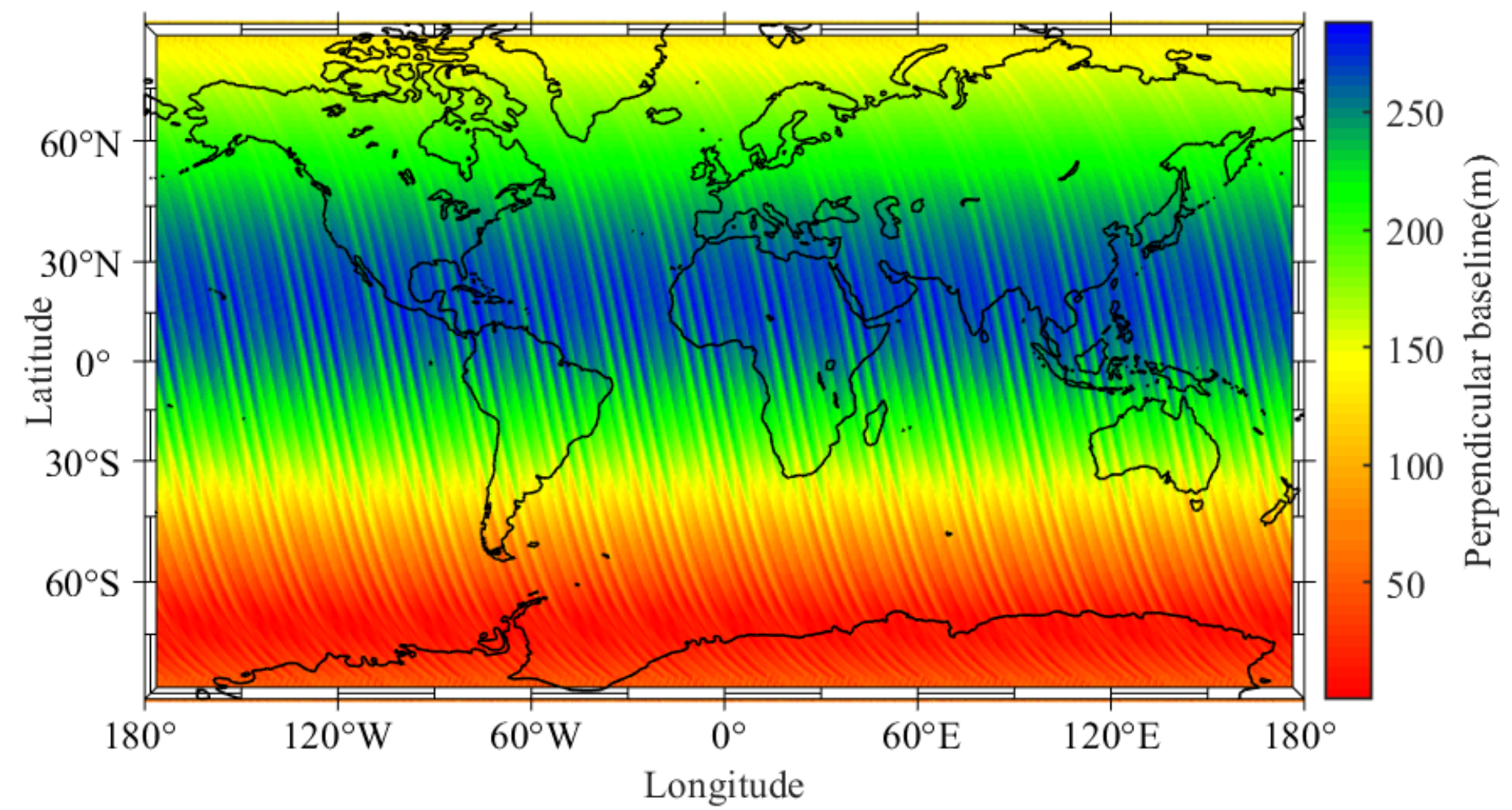}
  \caption{}
\end{subfigure}
\begin{subfigure}[b]{.48\textwidth}
  \includegraphics[width=\textwidth]{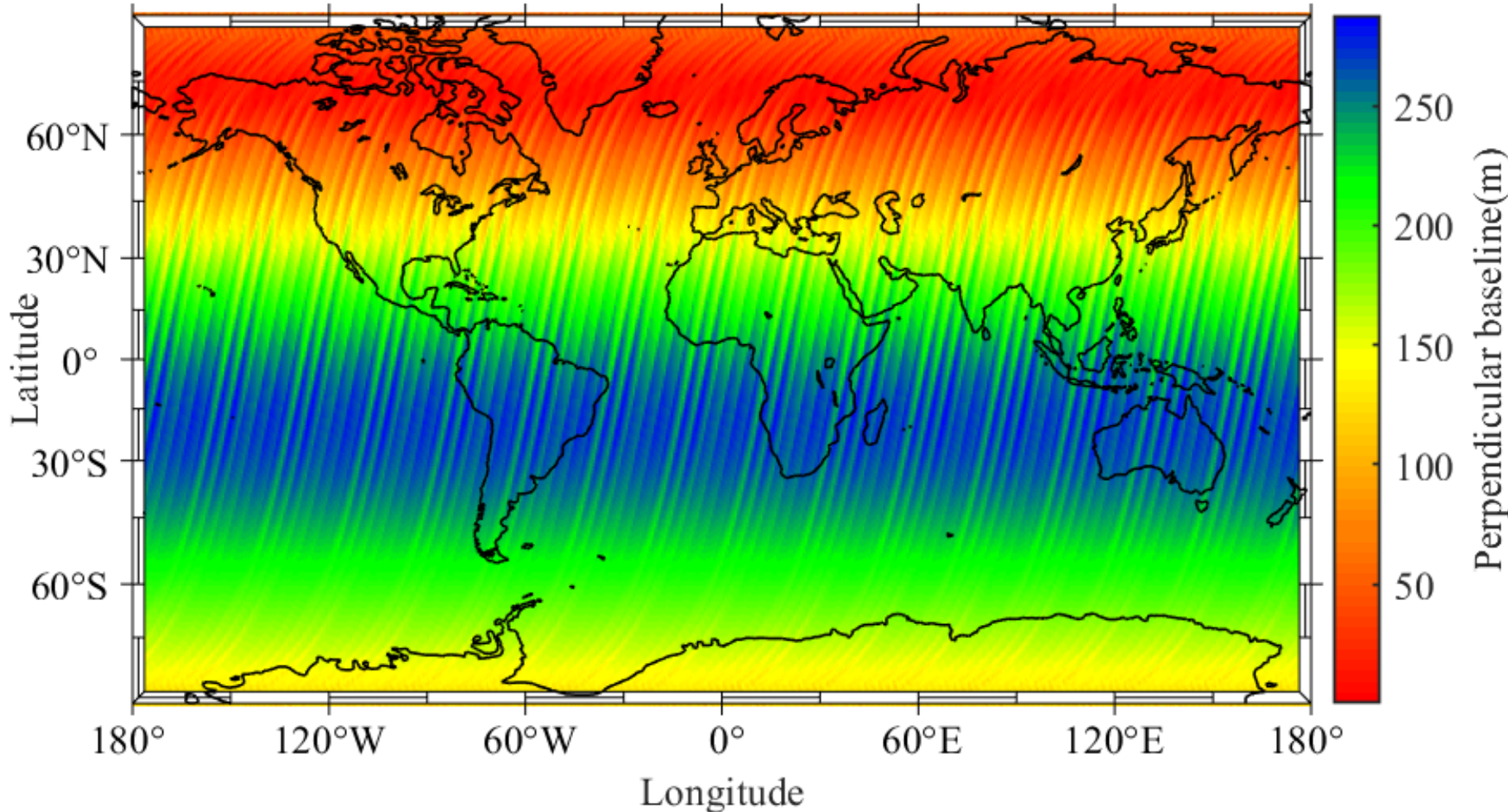}
  \caption{}
\end{subfigure} 
     \caption{\color{black} Variation of the perpendicular baseline. (a) Ascending track (b) Descending track. }
       \label{fig:baseline_variation}   
\end{figure}

{\color{black}
\subsection{Orbit configuration and Formation}
An essential design of the TDA-InSAR is the simultaneous acquisition of the two satellites, giving a long baseline dual-satellite interferogram. This operational mission requires a coordinated formation of two satellites flying in similar orbit. 
In the practical application, the helix orbit adopted by TanDEM~\cite{Krieger2007,Alberto2003} has been proven to be a reliable formation, which can be used in the TDA-InSAR design. A simple sketch of the helix formation is show in Fig.~\ref{fig:helix}. 
The horizontal displacement depends on the ascending node while the vertical displacement is related to the difference of the eccentricity. 

The relative displacement between the two satellites varies with the satellite altitude. Fig.~\ref{fig:baseline_variation} shows that the perpendicular baseline varies with the latitude, leading to different height performance. Additionally, this simulation also indicates that the perpendicular baseline of the ascending data in northern latitudes is longer than the in southern latitudes while the performance of the descending data is better in southern latitudes. 
According to the performance analysis in Fig.~\ref{fig:mode} (c), the optimal baseline of the configuration 2 ranges from 50 m to 300 m with an antenna baseline of 15 m. 
The orbit formation is able to give an optimal baseline ranging from 150 m to 300 m, which satisfied the requirement of the optimal baseline design.

}

\section{Simulation-Based Performance Evaluation}
In this section, simulation based analysis is conducted to evaluate the performance of the proposed TDA-InSAR. Two typical scenarios, i.e. built-up objects and vegetation canopies, are tested and followed by the analysis of the main error sources. 

\subsection{Performance of 3D Reconstruction}

\subsubsection{Built-up objects} 
\label{sec:target}

\begin{figure}[htbp]
     \centering
     \includegraphics[width=0.45\textwidth]{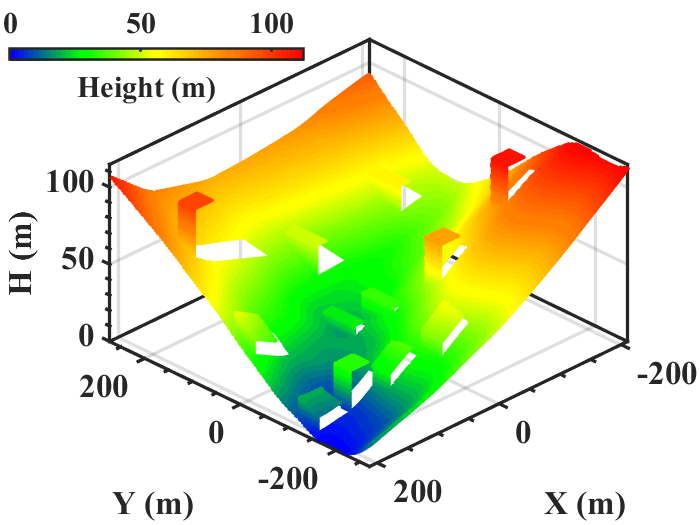}
     \caption{ Scene data for the simulation. }
     \label{fig:dem}    
\end{figure}  

\begin{figure*}[htbp]
\centering
\begin{subfigure}[b]{.32\textwidth}
  \includegraphics[width=\textwidth]{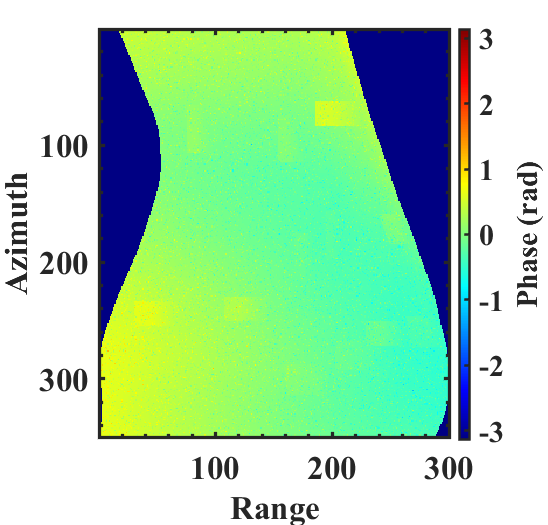}
  \caption{}
\end{subfigure}
\begin{subfigure}[b]{.32\textwidth}
  \includegraphics[width=\textwidth]{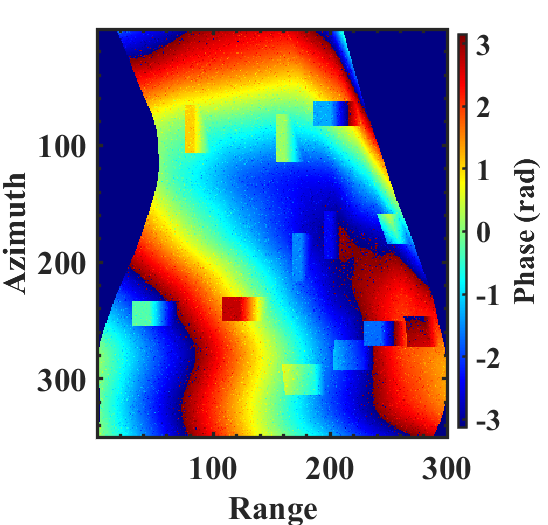}
  \caption{}
\end{subfigure}
\begin{subfigure}[b]{.32\textwidth}
  \includegraphics[width=\textwidth]{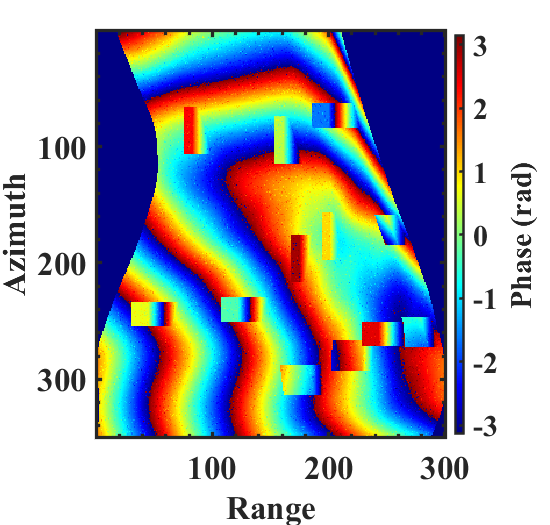}
  \caption{}
\end{subfigure} 
     \caption{ Simulated MB interferograms with the baselines of (a) 15 m (b) 150 m and (c) 300 m. }
       \label{fig:ifg}   
\end{figure*}

\begin{figure*}[tbp]
\centering
\begin{subfigure}[b]{.3\textwidth}
  \includegraphics[width=\textwidth]{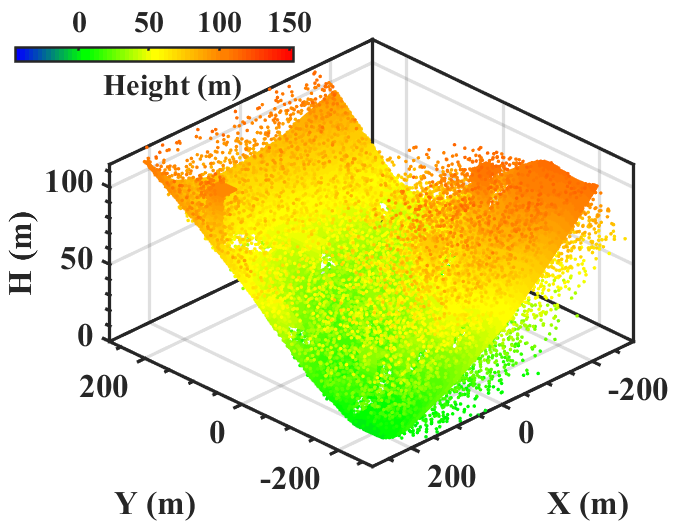}
  \caption{}
\end{subfigure}
\begin{subfigure}[b]{.3\textwidth}
  \includegraphics[width=\textwidth]{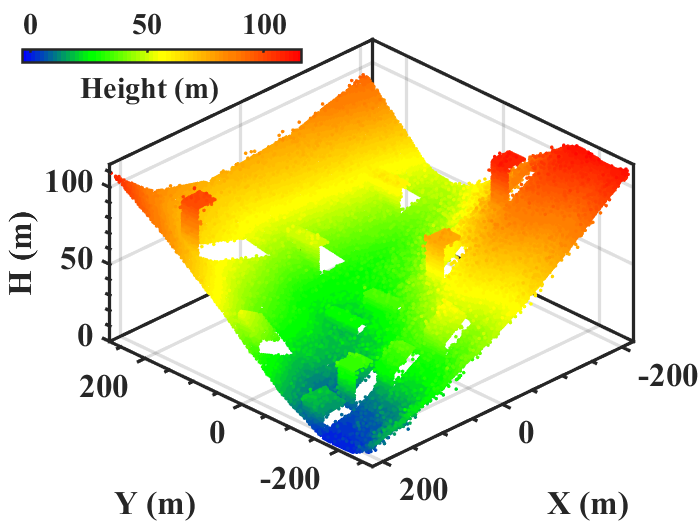}
  \caption{}
\end{subfigure}
\begin{subfigure}[b]{.3\textwidth}
  \includegraphics[width=\textwidth]{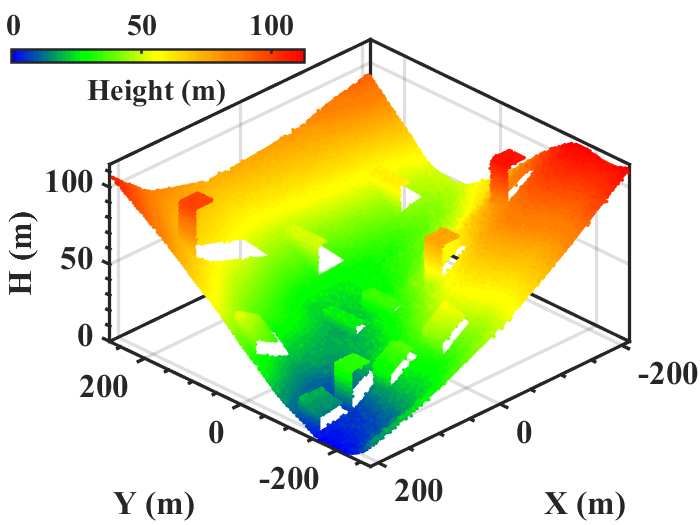}
  \caption{}
\end{subfigure} 
\begin{subfigure}[b]{.32\textwidth}
  \includegraphics[width=\textwidth]{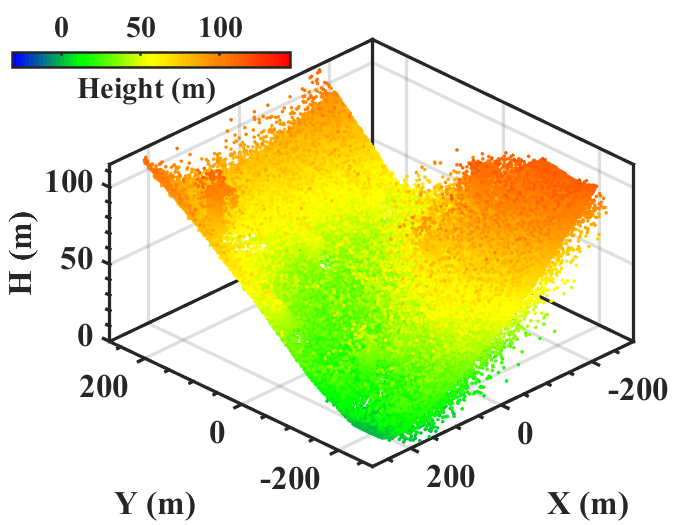}
  \caption{}
\end{subfigure}
\begin{subfigure}[b]{.32\textwidth}
  \includegraphics[width=\textwidth]{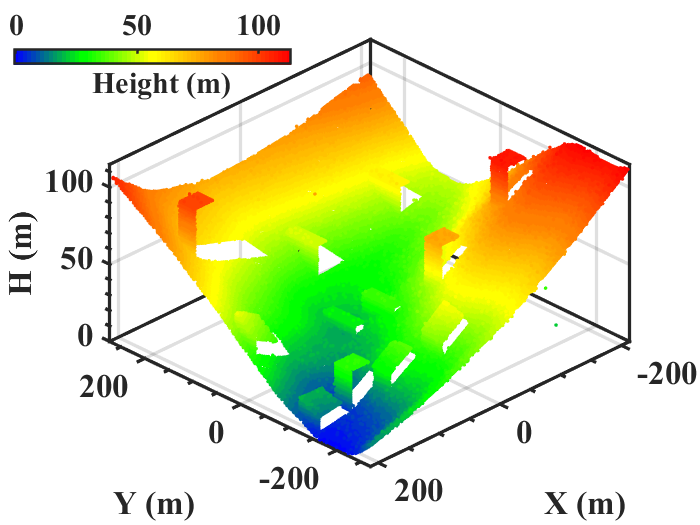}
  \caption{}
\end{subfigure}
\begin{subfigure}[b]{.32\textwidth}
  \includegraphics[width=\textwidth]{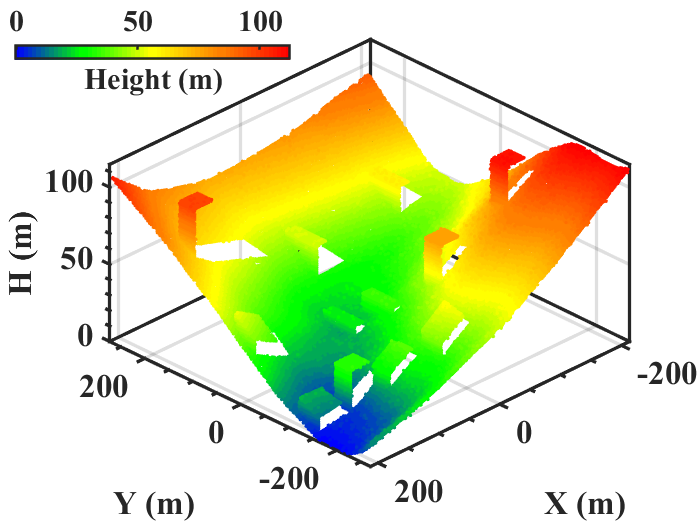}
  \caption{}
\end{subfigure}
\begin{subfigure}[b]{.32\textwidth}
  \includegraphics[width=\textwidth]{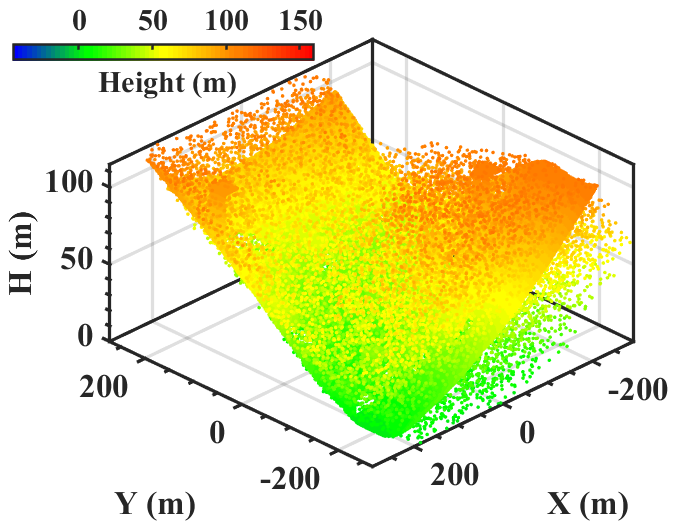}
  \caption{}
\end{subfigure}
\begin{subfigure}[b]{.32\textwidth}
  \includegraphics[width=\textwidth]{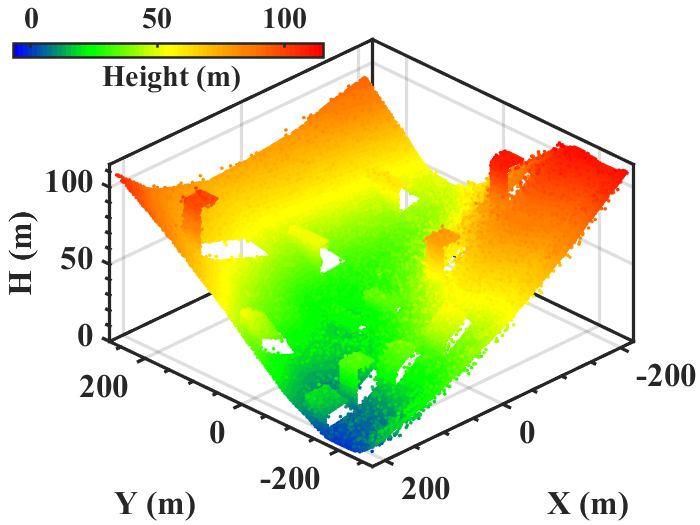}
  \caption{}
\end{subfigure}
\begin{subfigure}[b]{.32\textwidth}
  \includegraphics[width=\textwidth]{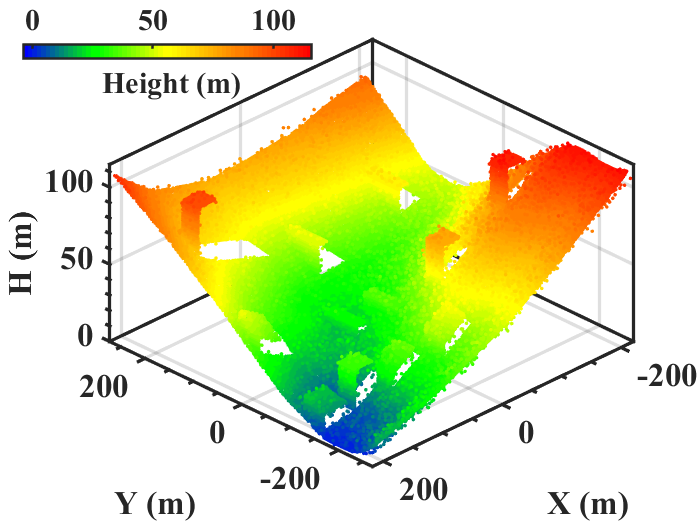}
  \caption{}
\end{subfigure} 
\begin{subfigure}[b]{.32\textwidth}
  \includegraphics[width=\textwidth]{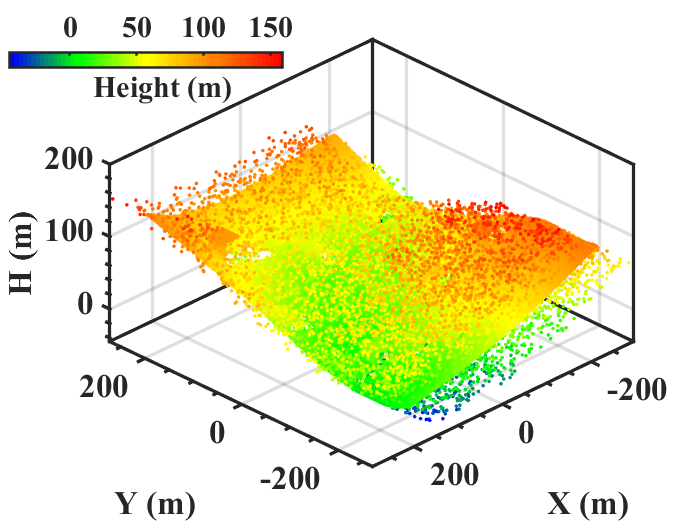}
  \caption{}
\end{subfigure}
\begin{subfigure}[b]{.32\textwidth}
  \includegraphics[width=\textwidth]{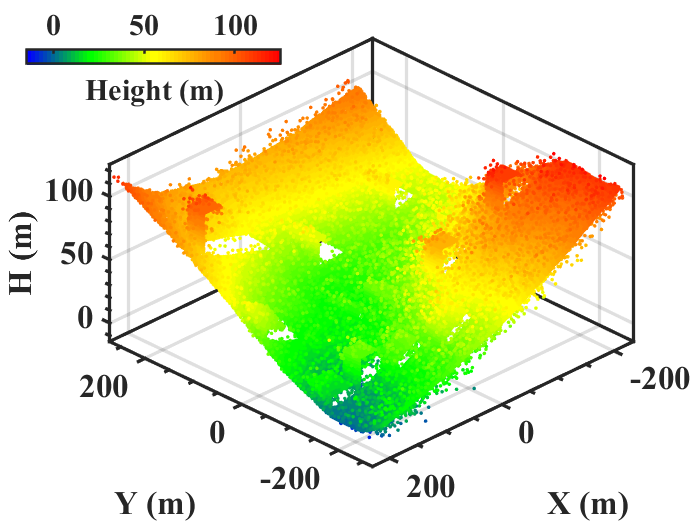}
  \caption{}
\end{subfigure}
\begin{subfigure}[b]{.32\textwidth}
  \includegraphics[width=\textwidth]{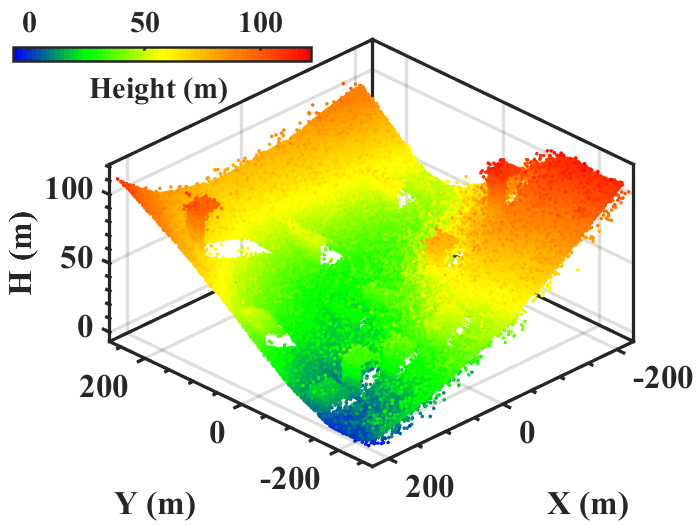}
  \caption{}
\end{subfigure}  
     \caption{ Investigation of 3D reconstruction. First column: result of SBI. Second column: result of MBI. Third column: result of LBI. Each row denotes the results by different baseline configurations. }
       \label{fig:simulation_mode}   
\end{figure*}

Based on a real elevation model and simulated ground targets, as shown in Fig.~\ref{fig:dem}, the MB interferograms are obtained using a multi-dimension coherent scattering model~\cite{Xue2020}. The real elevation model covers a range of 400 m $\times$ 500 m with a maximal height difference of 100 m and 12 ground targets are added. Using a radar wavelength of 31 mm, the corresponding interferograms with the perpendicular baselines of 15, 150 and 300 m are obtained, respectively, as shown in Fig.\ref{fig:ifg}. It shows that the phase change in the shortest baseline interferogram is very slow, which is good for spatial PU but poor for height inversion. The long baseline interferogram can not be unwrapped using the spatial PU due to the rapid phase change induced by the ground targets. 

According to the performance analysis in Fig.\ref{fig:mode}, the optimal baseline design of the tandem dual-antenna SAR varies with the baseline configuration. 
Assuming the coherence of 0.99 and antenna baseline 15 m, the optimal satellite baselines with different TDA-InSAR baseline configurations are 200, 300, 150 and 100 m, respectively. 
Then 3D reconstruction using asymptotic 3D PU are obtained, as shown in Fig.~\ref{fig:simulation_mode}. The corresponding precision of the estimated relative height is shown in Table.~\ref{tab:2}.
According to Figs.~\ref{fig:simulation_mode} (c) (f) (i) and (l), the final height precision only depends on the baseline length of LBI. So the 3D reconstruction obtained by the baseline configuration 2 shows the best performance.


\begin{table} 
\caption{Relative height precision by different TDA-InSAR baseline configuration}
\begin{tabularx}{\linewidth}{lXXXX}
\toprule
{Baseline configuration} & 1 & 2 & 3 & 4\\
\midrule
SBI & 8.61 & 6.45 & 14.04 & 10.68 \\
MBI & 0.92 & 0.48 & 1.11 & 2.55 \\
LBI & 0.41 & 0.31 & 1.08 & 2.27\\
\bottomrule
\end{tabularx}
\label{tab:2}
\end{table}

\subsubsection{Vegetation canopies}

\begin{figure}[htbp]
     \centering
     \includegraphics[width=0.45\textwidth]{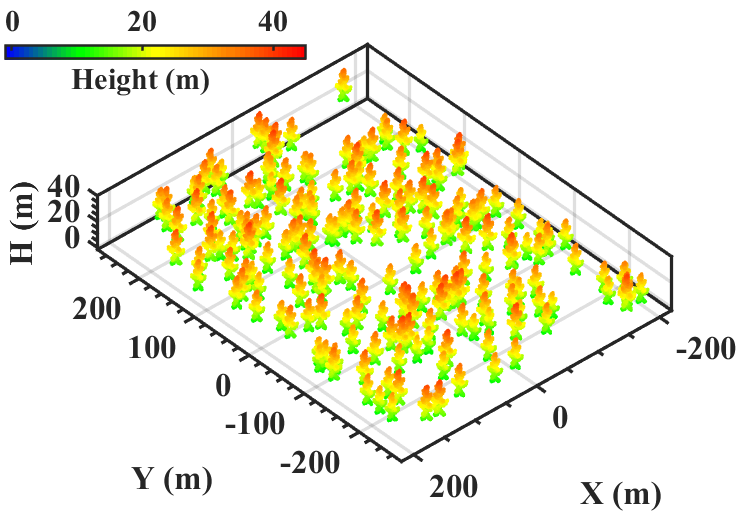}
     \caption{ Simulated 3D forest.}
     \label{fig:tree}    
\end{figure}  

\begin{figure}[htbp]
\centering
\begin{subfigure}[b]{.43\textwidth}
  \includegraphics[width=\textwidth]{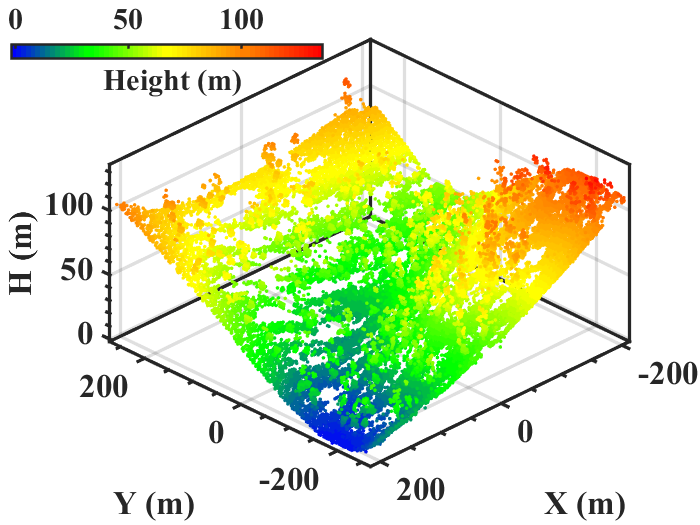}
  \caption{}
\end{subfigure}
\begin{subfigure}[b]{.43\textwidth}
  \includegraphics[width=\textwidth]{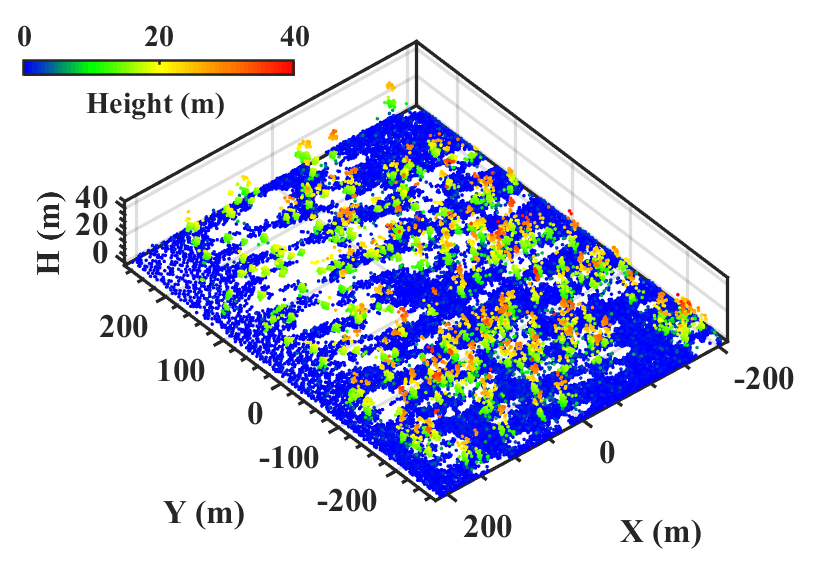}
  \caption{}
\end{subfigure}
     \caption{ 3D forest reconstruction (a) estimated heights of the radar scatterers (b) forest height inversion without the ground elevation change. }
       \label{fig:tree_hgt}   
\end{figure}

Virtual 3D scenes with deterministic fractal trees on DEM are generated using the following steps. First, the trees are parsed into canonical components or scatterers (i.e., dielectric cylinders and disks). Then, the generalized Rayleigh-Gans (GRG) approximation and the infinite cylinder approximations are used for coherent scattering calculations~\cite{Karam1988a}. 
Four-path multiple scattering mechanisms, i.e., direct scatterer scattering, ground-scatterer scattering, scatterer-ground scattering, and scatterer-ground-scatterer scattering, between vegetation and the ground are considered,~\cite{Xue2020}. Additionally, the scattering matrix of the rough surface under the vegetation can be obtained by the small perturbation method~\cite{Williams2006coherent,Jinyaqiu2013Polarimetric}.  
Assuming Foldy’s approximation~\cite{Tsang1985}, the extinction induced by electromagnetic waves penetrating the scatterers in the vegetation is also considered. 
In this study, a total of 200 trees with an average height of 30 m are simulated and they are randomly distributed in the whole area, which is shown in Fig.~\ref{fig:tree}. Note that the radar with a long wavelength is usually used in the practical tree height measurement to get a good penetration. In this simulation, the radar wavelength is setting to 0.24 m and the optimal baselines are also revised accordingly.

Using the same procedure in Section~\ref{sec:target}, the estimated heights of the radar scatterers is obtained. Since the estimated height in InSAR approach is defined with respect to a specified reference point, the final absolute height contains a constant height offset, leading to significant coordinate offsets in both horizontal and vertical directions. Such height offset can be estimated using a searching strategy using an initial reference DEM. Details can be found in~\cite{Hu2019railway}. The estimated height of the radar scatterers after absolute height correction is shown in Fig.~\ref{fig:tree_hgt} (a). To show the inversed forest height clearly, the initial DEM is used to remove the ground elevation change, as shown in Fig.~\ref{fig:tree_hgt} (b). 

Regarding the quantified analysis, three indicators, i.e., coverage, mean error (ME) and root mean square error (RMSE) are used to evaluate the reconstructed 3D forest height. 
Since the image resolution is lower than the simulated trees, the complex value of every pixel is the summation of all elementary scatterers within a resolution cell, which increases the difficulties in associating the radar scatterers to the corresponding trees. In this study, the nearest neighbor search approach is used to snap the point cloud in the simulated trees to their most likely radar scatterers. Based on the searching result, the ME, RMSE and coverage of the reconstructed 3D forest height are 0.14 m, 1.78 m and 59.7, correspondingly. 

Similar 3D forest reconstruction obtained by conventional 3D PU in \cite{Hu2022tribeam} shows a height RMSE of 1.25 m using 29 repeat-pass interferograms. To further evaluate the similarity between the forest height obtained by conventional and asymptotic 3D PU statistically, the two-sample F test is formulated as follows~\cite{DeGroot2012}
\begin{equation}
F_0 = \frac{\sigma_1^2}{\sigma_2^2}
\end{equation}
where $\sigma^2$ denotes the sample variance. Note that the impacts of both atmospheric delay and orbit error are neglected in this part, which will be investigated in the following section. The main factors that affect the precision of the estimated tree height are system noise. 
Then the statistical value $F_0$ can be calculated with the assumption that the height error follows a Gaussian distribution. 
Setting the significance level to $\alpha = 0.02$, the critical value $F_\alpha$ is 4.5. In this case, $F_0$ equals to 1.99, which is smaller than the critical value $F_\alpha$, showing that there are no significant differences on the estimated relative height by either conventional or asymptotic 3D PU. 
So the MB interferograms obtained by the TDA-InSAR can achieve fast 3D forest reconstruction.

\subsection{Impact of the main error sources}

\begin{figure}
\begin{subfigure}[b]{.4\textwidth}
  \includegraphics[width=\textwidth]{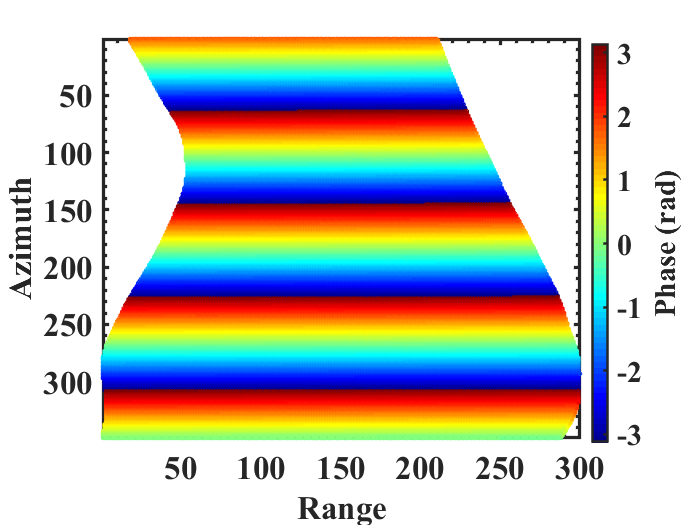}
  \caption{}
\end{subfigure} 
     \begin{subfigure}[b]{.4\textwidth}
  \includegraphics[width=\textwidth]{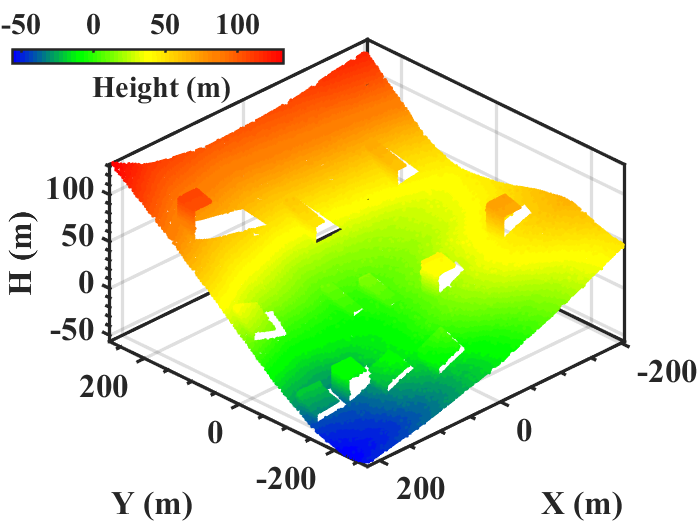}
  \caption{}
\end{subfigure} 
\caption{ Impact of the orbit error. (a) simulated orbit error (b) height estimation with the orbit error. }
     \label{fig:orb}  
\end{figure} 
\subsubsection{Orbit error}
Both mono-static and bi-static dual-satellite interferograms will suffer from the orbit error. In this demonstration, the parameters of the baseline configuration 2 was used to get the simulated MB interferograms and the orbit error was simulated using the nonlinear model in~Eq.(\ref{eq:orb_model}). 
The parameters $\delta B_c$ and $\delta B_n$ are set to 0.3 and 0.1 m while $\delta B_c$ and $\delta B_n$ are set to 0.02 and 0.02 m/PRF, respectively. The corresponding simulated orbit error phase is shown in Fig.~\ref{fig:orb} (a). 
It is obvious that the orbit error is spatially correlated signal, which will be mixed with the estimated height during the spatial PU. Fortunately, the dual-antenna interferogram is independent of the orbit error, which can be a good initialization for the spatial PU of the tandem satellite interferogram. 
The result of asymptotic 3D PU with orbit error is shown in Fig.~\ref{fig:orb} (b), showing that the orbit error will significantly bias the final height estimation.
The standard deviation of the height difference between the estimated height and the true value is 26.53 m. 

\begin{figure}[htbp]
\centering
\begin{subfigure}[b]{.45\textwidth}
  \includegraphics[width=\textwidth]{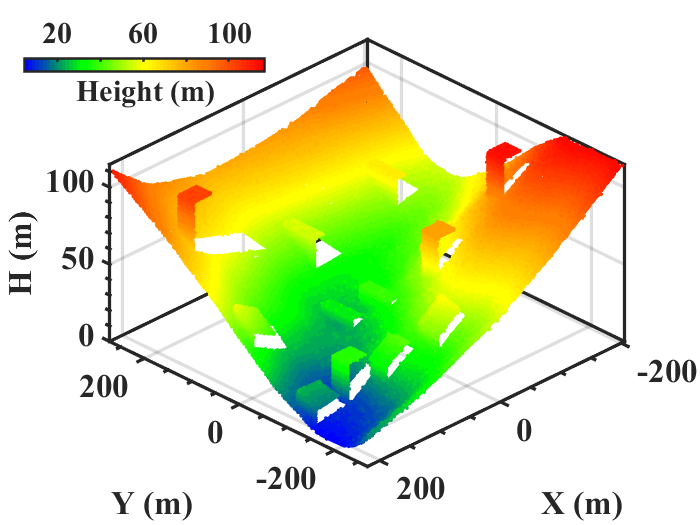}
  \caption{}
\end{subfigure}
\begin{subfigure}[b]{.45\textwidth}
  \includegraphics[width=\textwidth]{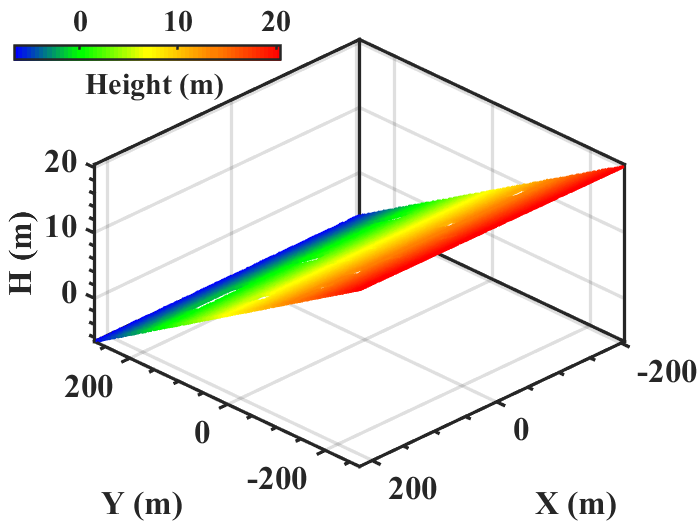}
  \caption{}
\end{subfigure}
     \caption{ Results of the height inversion with the orbit error compensation. (a) height estimation with the orbit error compensation (b) estimated orbit error. }
       \label{fig:orbit_correct}
\end{figure}

Using the proposed height inversion with the orbit error compensation, the final height estimation and the estimated orbit error are shown in Figs.~\ref{fig:orbit_correct} (a) and (b). 
The standard deviation of the height difference between the refined height and the true value is 0.33 m, which is consistent with the relative height precision by the simulation without the orbit error, see Table.~\ref{tab:2}.


\subsubsection{Atmospheric delay in bi-static interferogram}
{\color{black}Eliminating the atmospheric effect is the main challenge for the spaceborne SAR interferometry due to its spatio-temporal variation. 
For baseline configurations 1, 2 and 3 of the TDA-InSAR system, the dual-satellite interferogram is acquired using a bi-static mode, so the radar wave would propagate through the same tropospheric refractivity but different path. 
The impact of the atmospheric effect caused by the path difference should be investigated.}

In fact, a large part of the total atmospheric delay is almost constant, i.e., the ionospheric and hydrostatic troposphere delay~\cite{leijen14}. The main contributor to the spatial variation is the turbulence, which is effectively due to the water vapor distribution. 
{\color{black}Using the numerical atmospheric model, we investigate the impact of turbulence on the height inversion. The model used in this research is DALES, the Dutch Atmospheric Large Eddy Simulation model~\cite{Heus2010}, which can provide reliable simulation of the 3D tropospheric refractivity distribution. 

In the practical process, the simulation was conducted based on radiosonde and ground observations in Oklahoma and Kansas, USA~\cite{brown2002largeddy}. Additionally, 
}the simulation is characterized by shallow cumulus convection with a cloud cover between 20 and 30 percent, which is the typical fair-weather clouds over continental mid-latitudes~\cite{Hahn2007}. Then we get a 3D reflectivity distribution with the relevant parameters shown in Table.~\ref{tab:3}.
{\color{black}With the 3D refractivity distribution, the tropospheric delay along the line of sight can be computed using ray-tracing~\cite{Landon2011}. Since the refractivity change due to the bending error for typical SAR incidence angles can be negligible ~\cite{Park1996}, }the total tropospheric delay for one acquisition is obtained by integrating the refractivity from the elevation of the target to the total height of the simulated troposphere. 





\begin{figure}[htbp]
\centering
\begin{subfigure}[b]{.35\textwidth}
  \includegraphics[width=\textwidth]{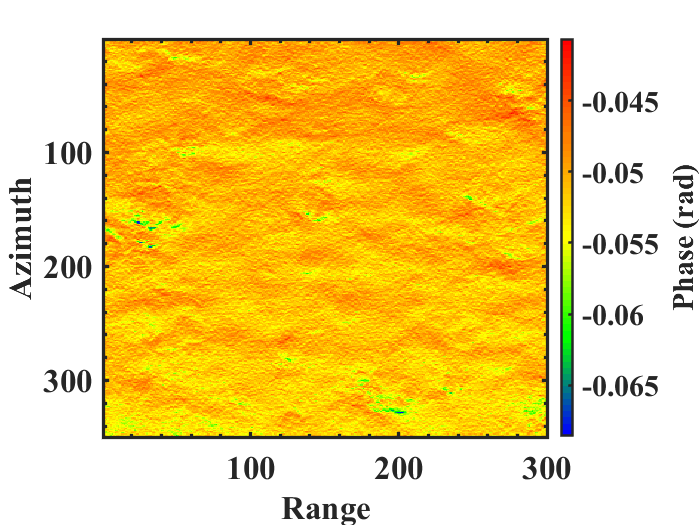}
  \caption{}
\end{subfigure}
\begin{subfigure}[b]{.35\textwidth}
  \includegraphics[width=\textwidth]{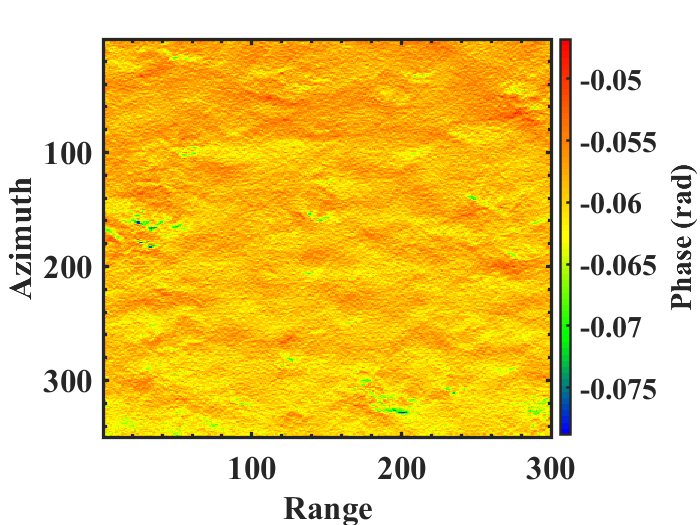}
  \caption{}
\end{subfigure}
\begin{subfigure}[b]{.35\textwidth}
  \includegraphics[width=\textwidth]{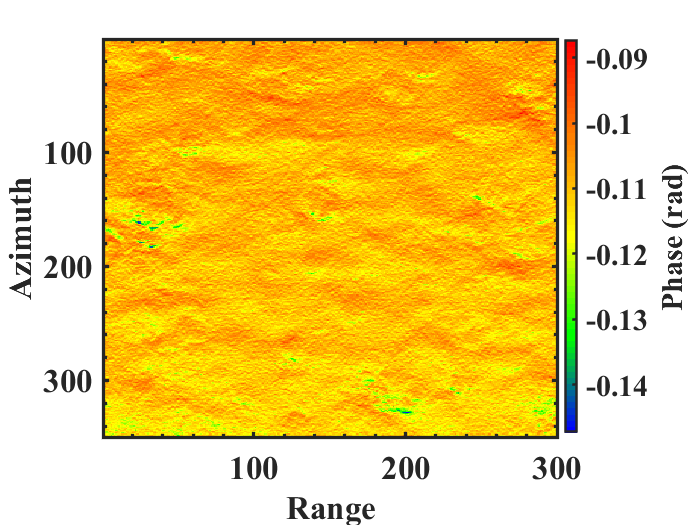}
  \caption{}
\end{subfigure} 
     \caption{ \color{black} Simulated synthetic tropospheric interferograms using the baseline configuration 2 with the perpendicular baselines of (a) 15 m (b) 150 m (c) 300 m  }
       \label{fig:atmos}   
\end{figure}

{\color{black}The evaluation of atmospheric effect due to the path difference was demonstrated using the parameters of the baseline configuration 2. Then the atmospheric interferograms, synthetic interferograms only sensitive to atmospheric delay variability with the perpendicular baselines of 15, 150 and 300 m are obtained, as shown in Fig.~\ref{fig:atmos}. 
Since the perpendicular baselines of the TDA-InSAR system} are very small compared with the radar altitude, the integration paths between different acquisitions are almost the same, leading to very limited phase change. 
Additionally, {\color{black}Fig.~\ref{fig:atmos} also shows that }such atmospheric delay will increase with the {\color{black}perpendicular} baseline, which may be confused with the estimated height. 

\begin{figure}[htbp]
     \centering
     \includegraphics[width=0.35\textwidth]{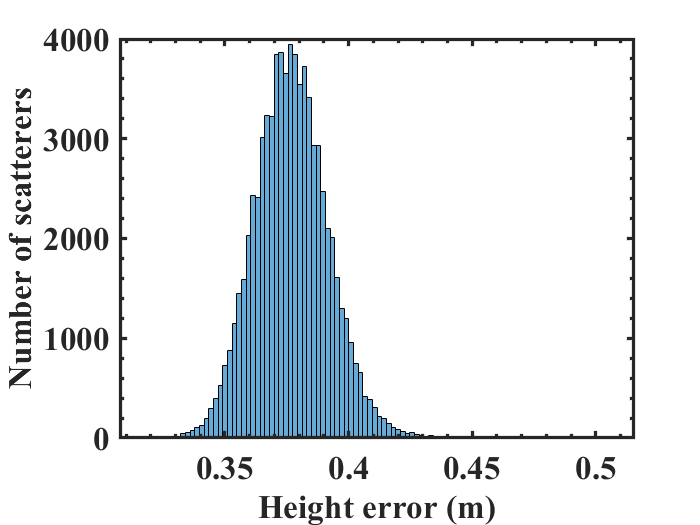}
     \caption{ Histogram of the height error induced by the atmospheric delay.}
     \label{fig:tro}    
\end{figure}

In this case, the histogram of the height bias {\color{black}caused by the atmospheric effect} is shown in Fig.~\ref{fig:tro}, {\color{black}indicating} that the atmospheric delay will lead to a significant height offset. Fortunately, such offset can be neglected since InSAR only gets the height estimation relative to a specified reference point. 
Moreover, the spatial variation of the height bias is only centimeter level, which is much smaller than the expected height precision. Thus, {\color{black}the atmospheric effect can be neglected if the TDA-InSAR works in a bi-static mode. }

\begin{table} 
\caption{Parameters of the refractivity distribution {in the simulation}}
\begin{tabularx}{\linewidth}{lX}
\toprule
{Parameters} & values \\
\midrule
Scale (km $\times$ km) & 49.3$\times$ 49.3 \\
Horizontal resolution (m$\times$m) & 40$\times$ 40 \\
Maximal height (m) & 4500 \\
Vertical resolution (m) & 40 \\
\bottomrule
\end{tabularx}
\label{tab:3}
\end{table}


\subsubsection{Atmospheric delay in mono-static interferogram}

\begin{figure}[htbp]
     \centering
     \includegraphics[width=0.4\textwidth]{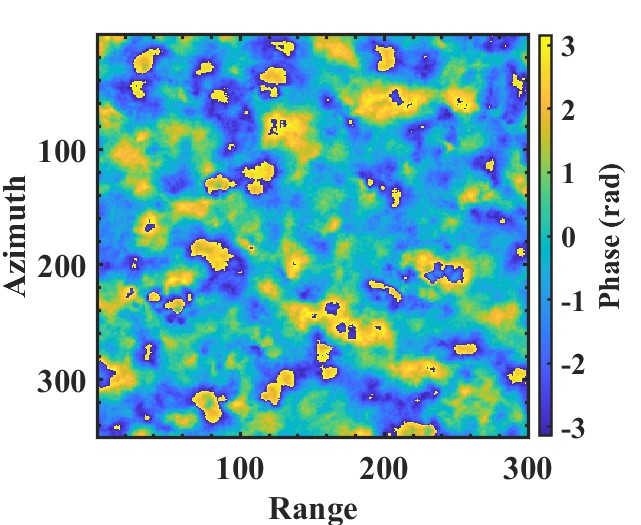}
     \caption{ Simulated synthetic tropospheric interferogram {\color{black}with a temporal baseline of 15 minute}.}
     \label{fig:atmos_ifg}    
\end{figure}

{\color{black}On the contrary, the baseline configuration 4 of the TDA-InSAR system doesn't require the strict synchronization. If the two satellites works separately in a mono-static mode, the change of the tropospheric refractivity will increase the variation of atmospheric effects. Based on the DALES model, we simulate two 3D tropospheric refractivity distributions with a time interval of 15 minute and get the}
corresponding synthetic tropospheric interferogram, as shown in Fig.~\ref{fig:atmos_ifg}. It shows that the phase change caused by the atmospheric effect exceed one cycle, equaling to the contribution of 75 meter height, which would significantly bias the final height estimation. 

\begin{figure}[htbp]
\centering
\begin{subfigure}[b]{.45\textwidth}
  \includegraphics[width=\textwidth]{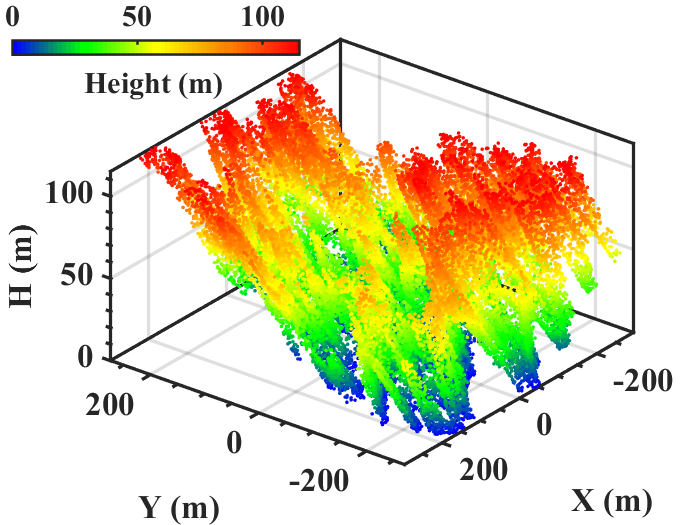}
  \caption{}
\end{subfigure}
\begin{subfigure}[b]{.4\textwidth}
  \includegraphics[width=\textwidth]{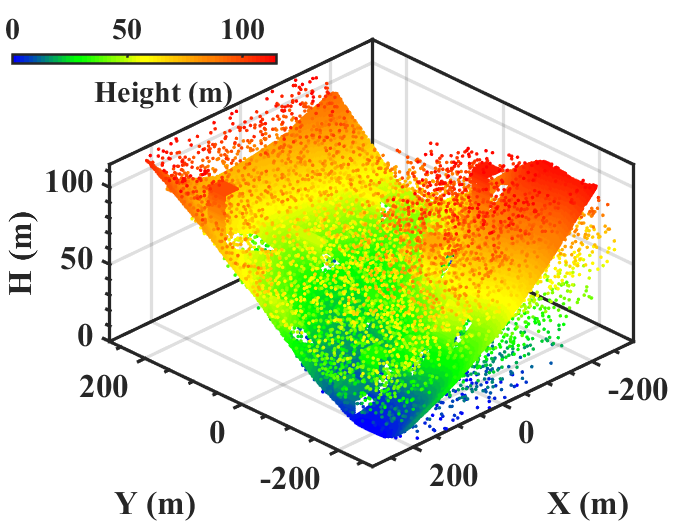}
  \caption{}
\end{subfigure}
     \caption{ Height estimation without (a) and with (b) atmospheric delay correction using the single dual-antenna satellite SAR.}
       \label{fig:atmos_mode4}   
\end{figure}

If the atmospheric delay is not considered during the process, the final height estimation is actually the atmospheric delay, as shown in Fig.~\ref{fig:atmos_mode4}(a). On the contrary, if {\color{black}additional parameters are used to model} the atmospheric delay during the process, the final height estimation is better, as shown in Fig.~\ref{fig:atmos_mode4}(b). However, compared Fig.~\ref{fig:simulation_mode} (l) with Fig.~\ref{fig:atmos_mode4} (b), 
{\color{black}height precision of the baseline configuration 4 is poorer than that of the baseline configuration 2 since the atmospheric effect will bias the ambiguity estimation. Therefore, strict synchronization is necessary to avoid the atmospheric effect, which can provide a good height precision.}

\section{Conclusions}
In this paper, we propose a new {\color{black}concept of} TDA-InSAR to acquire the specified MB interferograms for asymptotic 3D PU, which enables a reliable 3D reconstruction using very sparse acquisitions. {\color{black}Two indicators, relative height precision and successful phase unwrapping rate, are used to develop the optimal baseline design and the performance of difference baseline configurations is evaluated correspondingly. Assuming that the antenna length is 15 m, the optimal satellite baseline of the bi-static mode can be selected in a flexible range from 50 m to 300 m while that of the mono-static mode is fixed at 100 m.} {\color{black}The flexible baseline selection will guarantee the optimal baseline design when the orbit control is inaccurate. Although the bi-static mode requires a strict signal synchronization between the tandem satellites, increasing the complexity of the hardware design, the use of longer baseline interferogram leads to a better relative height precision. }

Additionally, simulation-based evaluation of one example configuration shows that the proposed system enables a 3D reconstruction with a relative height precision of 0.3 m in built-up or man-made objects and that of 1.7 m in vegetation canopies. {\color{black}This single-pass SAR system }combines the advantages of both multi-antenna and multi-satellite SAR and thus shows a good {coherence in \color{black}both built-up objects and vegetation canopies}. 
{\color{black}
Considering the impact of the atmospheric delay, the LES model is used to get a realistic atmospheric simulation and the analysis of the atmospheric effects show the slight path difference due to the baseline diversity will not affect the height inversion but the temporal refractivity change leads to significant height bias. It indicates that the atmospheric effect can be neglected if the proposed TDA-InSAR works in a bi-static mode. 
Using the asymptotic 3D PU, the dual-satellite interferogram can be successfully unwrapped with the help of the dual-antenna interferogram and the orbit phase trend can be well compensated. 
Thus, the proposed TDA-InSAR system enables} the fast 3D reconstruction {\color{black}in a single flight}, showing its scientific importance in many applications, such as mapping terrain, target recognition or forest height inversion, especially for users demanding a single-pass acquisition. 

\section*{Declaration of competing interest}
The authors declare that they have no known competing financial interests ot personal relationships that could have appeared to influence the work reported in this paper.

\section*{Acknowledgments}
This work was supported in part by the National Nature Science Foundation of China under Grant 61991422 and 62201158.

\printcredits

\bibliographystyle{cas-model2-names}

\bibliography{insarref}



\end{document}